\PassOptionsToPackage{table,xcdraw}{xcolor}
\documentclass[a4paper,fleqn]{cas-sc}

\usepackage[numbers]{natbib}
\def\tsc#1{\csdef{#1}{\textsc{\lowercase{#1}}\xspace}}
\tsc{WGM}
\tsc{QE}
\usepackage[utf8]{inputenc}
\usepackage[T1]{fontenc}
\usepackage{etoolbox}
\usepackage{siunitx,comment}
\usepackage[normalem]{ulem}
\usepackage{lettrine}
\usepackage{url}
\usepackage{graphicx}
\usepackage{dcolumn}
\usepackage{bm}
\usepackage{amsmath,amssymb}
\usepackage{subcaption}
\usepackage{caption}
\usepackage{booktabs} % For professional tables
\usepackage{hyperref} % For clickable links
\usepackage{fancyhdr} % For custom headers and footers
\usepackage{svg}
\usepackage{cleveref} % For improved referencing
\usepackage{standalone}
\usepackage{tikz}
\usetikzlibrary{arrows.meta, positioning}
\usepackage{pifont}
\usepackage{float}

\definecolor{plasmacol}{RGB}{253,178,134}
\definecolor{bluevec}{RGB}{100,160,215}
\definecolor{redvec}{RGB}{200,20,20}
\newcommand{\dV}{\:\mathrm{d}\Omega}
\newcommand{\dS}{\:\mathrm{d}S}

\newcommand{\curl}{\nabla \times }

\newcommand{\id}{{\rm I\! I}}

\crefname{equation}{Eq.}{Eqs.}
\crefname{figure}{Fig.}{Figs.}
\crefname{table}{Table}{Tables}
\crefname{section}{Section}{Sections}
\crefname{appendix}{Appendix}{Appendices}

\RenewDocumentEnvironment{Abstract}{o}
  {%
    \par\begingroup
    \setlength{\leftskip}{0pt}
    \setlength{\rightskip}{0pt}
    \setlength{\parindent}{0pt}
    \noindent\abstractname\par
    \vspace{-4pt}
    \noindent\rule{\linewidth}{0.2pt}\par
    \footnotesize
    \ignorespaces
    \everypar{\parindent=1.5em}
  }
  {\par\endgroup}

\expandafter\def\csname __first_footerline:\endcsname{%
  \begingroup
  \small\sffamily
  \csname __short_authors:\endcsname
  \endgroup
}

\begin{document}
\let\WriteBookmarks\relax
\def\floatpagepagefraction{1}
\def\textpagefraction{.001}

\shorttitle{On Spurious Mesh-Scale Oscillations Associated with Resonance-Cone Regimes in Wave-Plasma Simulations}    

\shortauthors{Hugo Bergerioux, Mario Merino}  

\title [mode = title]{On Spurious Mesh-Scale Oscillations Associated with Resonance-Cone Regimes in Wave-Plasma Simulations}

\author[1]{Hugo Bergerioux}[orcid=0000-0002-4316-4641]%[<options>]

\cormark[1]

\ead{hbergeri@ing.uc3m.es}

\affiliation[1]{organization={Department of Aerospace Engineering, Universidad Carlos III de Madrid},
            addressline={Av. de la Universidad, 30}, 
            city={Leganés},
            postcode={28911}, 
            state={Madrid},
            country={Spain}}

\author[1]{Mario Merino}[orcid=0000-0001-7649-3663]%[]

% Email id of the second author
\ead{mario.merino@uc3m.es}

\cortext[1]{Corresponding author}

\begin{abstract}
Spurious mesh-scale oscillations can compromise frequency-domain finite-element simulations of electromagnetic wave propagation in magnetized cold plasmas. We investigate their origin and mitigation through a verification framework combining a mixed Nédélec–Lagrange solver, a waveport accommodating non-orthogonal modes in anisotropic media, and an analytical generalized Fresnel benchmark at a vacuum–plasma interface.
Comparisons with this benchmark show that the observed oscillations occur in resonance-cone regimes and are absent from the analytical fields, confirming their numerical origin. Systematic tests assess the effects of approximation order, variational formulation, mesh geometry, and numerical collisionality. 
Plane-wave residual and Fourier spectral analyses suggest that local mesh variations induce spectral coupling to additional wavevector components, predominantly populating the dispersion branches in collisionless resonance-cone regimes where the reduced scalar equation loses ellipticity. 
Uniform rectangular grids recover the analytical wave structure, further supporting the role of mesh geometry and symmetry. However, strict symmetry requirements limit the applicability of this approach, motivating numerical regularization.
In resonance-cone regimes, introducing sufficient numerical collisionality suppresses the mesh-scale oscillations, yielding smooth electric-field solutions and reducing the error relative to the analytical benchmark. For one resonance-cone benchmark configuration, the error-minimizing damping decreases with mesh refinement according to an empirical $Z_{\mathrm n}\propto h^{1.48}$ scaling, sustaining error reduction consistent with second-order convergence and avoiding the accuracy limit imposed by fixed collisionality.
The framework is then applied to a waveguide-fed electron cyclotron resonance plasma source with non-uniform plasma and physical collisions. A joint mesh and damping study identifies an absorbed-power plateau for normalized numerical collision frequencies between $3\times10^{-3}$ and $3\times10^{-2}$, with differences across the tested meshes below $4\%$ throughout this range. These results provide configuration-dependent guidance for selecting numerical collisionality.
\end{abstract}

\maketitle

% ------------------------------------------------------
\section{Introduction}
Finite-element simulations of electromagnetic wave propagation and absorption in magnetized plasmas can exhibit spurious mesh-scale oscillations in strongly anisotropic propagation regimes.
A common approach employs full-wave, frequency-domain electromagnetic simulations based on the cold plasma approximation, in which the plasma is represented as an anisotropic dielectric medium governed by a complex permittivity tensor.
This wave solver can also be coupled with a transport code to compute self-consistent plasma and field distributions \cite{domi19a, svil21a}.
In contrast with time-marching electromagnetic particle-in-cell simulations \cite{porto2023, meri26_preprint}, this approach directly computes the time-harmonic response at the excitation frequency for a given plasma background, generally at a lower computational cost.
These simplifications make it attractive for parametric studies, although finite-temperature and nonlinear effects in the wave response are neglected.
An illustrative application of this simulation scheme is the modeling of Electrodeless Plasma Thrusters (EPTs) \cite{svil21a,jime22a}, a class of electric propulsion devices that use externally applied electromagnetic waves to heat electrons and sustain propellant ionization \cite{bathgate17, taka19}, and accelerate the resulting quasineutral plasma using a magnetic nozzle \cite{ahed10f,meri16g}, thereby eliminating the need for neutralizers or exposed electrodes.
In particular, Electron Cyclotron Resonance Thrusters (ECRTs) achieve this via resonant microwave absorption, typically through waveguide \cite{peterschmitt_impact, inch23a, inchingolo24} or coaxial \cite{jarr13b, desangles23} coupling.
For these devices, reliable predictions of power deposition are essential for understanding their operation, guiding design optimization, and assessing thruster performance. 

Early Finite-Difference (FD) studies of wave-plasma interaction used time-domain schemes \cite{young1994,cummer1997}. Two-dimensional, axisymmetric frequency-domain models were then developed for helicon sources, accounting for radial and axial plasma non-uniformity in cylindrical geometries \cite{mouz98,chen06b}.
These models were extended to include the divergent magnetic field and expanding near plume of Helicon Plasma Thrusters (HPTs) \cite{tian18a}, and subsequently the far plume and downstream Electron Cyclotron Resonance (ECR) surface \cite{jime22a}.
The Finite Element Method (FEM) on unstructured meshes provides an easier and more flexible approach to tackle complex geometric boundaries and material interfaces.
Early ECR plasma simulations by Yasaka et al. \cite{yasaka92} combined an azimuthal Fourier expansion with a two-dimensional Galerkin FEM wave solver. Three-dimensional FEM models were subsequently developed for ECR ion sources \cite{torrisi2014,masc15}, including mesh refinement around the resonance layers.
Regarding EPT applications, Sánchez-Villar et al. developed a mixed FEM wave solver coupled to plasma transport models for coaxial ECRTs \cite{svil21a,svil23}; local refinement reduced spurious oscillations near the upper-hybrid resonance in their simulations. 
Jiménez et al. \cite{jime23a} extended this mixed FEM formulation from $m=0$ to arbitrary azimuthal modes and applied it to an HPT. Their coupled wave-transport simulations linked downstream ECR absorption to an electron temperature maximum also observed experimentally. Compared with their earlier FD implementation, the FEM solver also reduced mesh sensitivity in the plume.

Numerical difficulties nevertheless remain in actual device simulations. Using the coupled wave-transport framework of Jiménez et al. \cite{jime23a}, Inchingolo et al. \cite{inch24b} simulated a waveguide-ECRT and obtained electric field maps dominated by short-wavelength oscillations over large portions of the plasma domain. 
They associated these patterns with resonance-cone regimes and identified the need for further convergence studies to assess possible spurious contributions.
Indeed, in ECRTs --- as in other devices --- non-uniform plasma density and magnetic-field profiles can produce extended regions supporting resonance cones \cite{jime22a,inch24b}. In these regimes, a branch of the homogeneous collisionless cold plasma dispersion relation admits an unbounded wavenumber as the angle between the wavevector and the background magnetic field approaches a critical value \cite{STIX92}. The model therefore supports arbitrarily short wavelengths, which can impose strong spatial-resolution requirements.
Related numerical difficulties have also been reported in fusion plasma simulations: Otin et al. reported spurious oscillations near the lower-hybrid resonance \cite{otin20}. Studies of resonance cones have also characterized their collisionless field singularities \cite{tier23} and shown that resolving their collisional field peaks can require very fine meshes \cite{tier24}. In the unstructured-mesh ECRT simulations considered here, rapid spatial oscillations persist over the tested range of mesh refinements.

Introducing additional collisionality is one approach to regularizing cold plasma wave calculations.
Effective collision frequencies have been used to approximate kinetic absorption in radio-frequency plasmas \cite{vahediAnalyticModelPower1995,gody99} and collisionless heating in ECR models \cite{williamson1992,yasaka92}.
Alongside these physical modeling considerations, Hagelaar et al. \cite{hage09} reported improved convergence in their FD ECR model with limited changes to the overall power deposition profile. In HPT simulations, Tian et al. \cite{tian18a} highlighted that finite collisionality limits the wavenumber near resonance cone directions, easing spatial resolution requirements, while Jiménez et al. \cite{jime22a} increased collisionality artificially to improve convergence but cautioned that excessive damping could obscure physical short-wavelength structures. Earlier finite-difference dispersion analyses also linked the suppression of spurious resonance cone branches to both collisionality and grid spacing \cite{meri17d}.
In cold plasma FEM simulations specifically, artificial collisionality has likewise been used to regularize resonance and cutoff layers \cite{migl23}. More recently, Otin et al. \cite{otin26} reported a collision-frequency threshold for obtaining stable edge-element solutions in an O--X mode-conversion benchmark.  

These studies motivate a verification-based assessment aimed at distinguishing physical short-wavelength fields from discretization-induced oscillations and selecting additional collisionality that suppresses numerical pollution without limiting accuracy under mesh refinement.
Following the presentation of the cold plasma model in \cref{SEC2_cold_plasma}, \cref{SEC3_discretization} introduces our numerical framework. The solver builds on earlier mixed finite-element models \cite{svil21a,jime23a} and is re-implemented and extended in Julia using the \texttt{Gridap.jl} FEM library \cite{gridap1,gridap2}, enabling comparisons between the standard wave and saddle-point formulations using first- and second-order mixed Nédélec--Lagrange elements.
The methodological contributions include a generalized modal waveport boundary condition for wave injection and absorption with non-orthogonal modes in anisotropic media, and an exact analytical verification benchmark based on a generalized Fresnel problem.
In \cref{SEC4_Fresnel}, this benchmark (a vacuum plane wave incident on a semi-infinite, homogeneous magnetized cold plasma) is described and provides reference fields across selected propagation regimes, including those supporting resonance cones.
Plane-wave residual and Fourier spectral analyses examine the role of spatial discretization, while convergence studies assess the effects of mesh geometry, approximation order, and variational formulation on both field accuracy and numerical pollution.
Using the collisionless Fresnel solution as a reference, we quantify the balance between suppression of spurious oscillations and damping-induced field changes, examining how the error-minimizing additional collisionality varies with mesh resolution.
In \cref{SEC5_ECRT}, the same waveport supports simulations of a waveguide-fed ECRT, adding numerical collisionality to the physical electron collision frequency.
We examine the sensitivity of power deposition to damping and mesh resolution, identifying a range of numerical collisionality over which the integrated absorbed power varies only weakly with both parameters, thus providing a practical criterion for selecting additional collisionality.

% ------------------------------------------------------

% ------------------------------------------------------
\section{Cold plasma wave model} \label{SEC2_cold_plasma}
The cold plasma model provides a simplified yet physically meaningful description of wave propagation in magnetized plasmas. It neglects thermal motion and describes the collective response of electrons and ions to electromagnetic fields, while retaining collisional momentum losses. The model is obtained from the linearization of the fluid momentum equation of each charged species for time-harmonic perturbations around a stationary equilibrium. 
The plasma response is described by a frequency-dependent anisotropic dielectric tensor $\bm{\kappa}$, following the standard cold plasma model \cite{STIX92,swan03,BITT04}.
In the absence of externally imposed volume sources, we seek time-harmonic solutions of Maxwell's equations of the form $\bm{E}(\bm{r},t)=\operatorname{Re}[\bm{\hat{E}}(\bm{r})\exp(-i\omega t)]$, where $\omega$ is the angular frequency and $\bm{\hat{E}}(\bm{r})$ is the complex electric-field phasor. Incorporating the plasma response through $\bm{\kappa}$, the frequency-domain wave equation takes the curl-curl form:
\begin{equation} \label{eqt-curl_curl_gen}
    \nabla \times \left(\nabla \times \bm{\hat{E}}\right)-\frac{\omega^{2}}{c^{2}} \bm{\kappa}\bm{\hat{E}} = \bm{0},
\end{equation}
where $c$ is the speed of light in vacuum, and the tensor $\bm{\kappa}$ is defined such that the electric displacement field becomes $\bm{D} = \bm{\varepsilon} \cdot \bm{E} = \varepsilon_0 \bm{\kappa} \cdot \bm{E}$. 
For a background magnetic field  aligned with the $z$-axis, $\bm{B}_0=B_0 \mathbf{\hat{e}}_z$ \cite{STIX92}, the cold plasma dielectric tensor is defined as:
\begin{equation} \label{eqt-dielectric_tensor}
    \bm{\kappa} = \left(\begin{array}{ccc}
    S & -i D & 0 \\
    i D & S & 0 \\
    0 & 0 & P
    \end{array}\right), \qquad \mathrm{with}\qquad S=\frac{R+L}{2}, \quad D=\frac{R-L}{2}. \\
\end{equation}
For other $\bm B_0$ field directions, the tensor components result from applying the appropriate rotation.
In the microwave-frequency range considered in this work, the inertial response of the ions is negligible compared to that of electrons, such that only the electron contribution is retained in the dielectric tensor elements.
The fundamental tensor components are then given by:
\begin{equation} \label{eqt:cold_plasma_RLP}
    R=1-\frac{\omega_{\mathrm{pe}}^{2}}{\omega\left[\left(\omega+i \nu\right)-\omega_{\mathrm{ce}}\right]}, \quad L=1- \frac{\omega_{\mathrm{pe}}^{2}}{\omega\left[\left(\omega+i \nu\right)+\omega_{\mathrm{ce}}\right]}, \quad \mathrm{and} \quad P=1- \frac{\omega_{\mathrm{pe}}^{2}}{\omega\left(\omega+i \nu\right)}.
\end{equation}
These parameters depend on the plasma frequency $\omega_{\mathrm{pe}} = \sqrt{n_e e^2/(m_\mathrm{e} \varepsilon_0)}$, the electron cyclotron frequency $\omega_{\mathrm{ce}} = eB_0/m_\mathrm{e}$, and an effective collisionality $\nu$, and they can be extracted from plasma transport simulation codes such as HYPHEN \cite{domi19a}, to compute the wave fields and the deposited power in the plasma.
Ultimately, the map of deposited power can be fed back to the transport code to establish a self-consistent electromagnetic-plasma transport simulation framework \cite{svil21a}.

\subsection{Plane-wave dispersion relation}

In an infinite homogeneous plasma, the wave equation admits plane-wave solutions for the electric field phasor of the form $\bm{\hat{E}}(\bm{r}) = \bm{\tilde{E}}\exp(i\bm{k}\cdot\bm{r})$, with $\bm{\tilde{E}}$ the constant complex amplitude vector and $\bm{k}$ the potentially complex wave vector. Applying the spatial Fourier transform $\nabla \times \bm{E} \rightarrow i \bm{k} \times \bm{E}$, the differential equation \cref{eqt-curl_curl_gen} reduces to an algebraic system:
\begin{align}
	\bm{k} \times \left(\bm{k} \times \bm{\tilde{E}}\right) + \frac{\omega^{2}}{c^{2}} \bm{\kappa} \bm{\tilde{E}} &= \bm{0}.
\end{align}
Introducing the refractive-index vector $\bm{n} = (c/\omega)\bm{k}$, we consider waves with $\bm{n} = n \cos(\theta) \mathbf{\hat{e}}_z + n \sin(\theta) \mathbf{\hat{e}}_x$, where $n$ may be complex and $\theta$ specifies a real propagation direction relative to the background magnetic field. The algebraic system can then be written using the dispersion tensor $\mathcal{L}(\bm{n})$:
\begin{align}
	\mathcal{L}(\bm{n})\bm{\tilde{E}} &= \left( \bm{n}\otimes\bm{n} - n^2 \id + \bm{\kappa} \right) \bm{\tilde{E}} = \bm{0},
\end{align}
where $\otimes$ denotes the dyadic product and $\id$ is the identity tensor.
For non-trivial wave solutions to exist, the determinant of the dispersion tensor must vanish:
\begin{align}
	\det\left(\mathcal{L}(\bm{n})\right) &= 0. \label{eqt-disp_rel}
\end{align}
Expanding this determinant yields the standard cold-plasma dispersion relation, commonly written in the Stix biquadratic form. For subsequent analysis, it is convenient to rearrange it into the Allis form:
\begin{align}
	\tan^{2}\theta &= -\frac{P\left(n^{2}-R\right)\left(n^{2}-L\right)}{\left(S n^{2}-R L\right)\left(n^{2}-P\right)}. \label{eqt-disp_rel_theta}
\end{align}
This dispersion relation describes wave propagation in the plasma and provides the conditions for cutoffs and resonances. In the collisionless model, cutoffs correspond to transitions where the refractive index vanishes and propagating modes become evanescent, whereas resonances arise when the refractive index diverges, $n\to\infty$. In practice, these resonant regimes are associated with strongly enhanced wave--plasma interaction, for which dissipative effects may lead to significant wave absorption.

The cutoff and resonance structure of the cold plasma model is commonly represented through the Clemmow--Mullaly--Allis (CMA) diagram, which classifies wave propagation regimes in the parameter space defined by the normalized frequencies $X = \omega_{\mathrm{pe}}^2/\omega^2$ (proportional to plasma density) and $Y = \omega_{\mathrm{ce}}/\omega$ (proportional to $B_0$). A third parameter when considering collisional problems is $Z = \nu/\omega$, the normalized collision frequency. The bounding curves divide the physical parameter space into distinct propagation regions. In the general cold plasma formulation, the CMA diagram comprises up to thirteen regions depending on the considered parameter range and species contributions \cite{STIX92, swan03, BITT04}. In the microwave-frequency regimes relevant to most low-temperature plasma sources, including ECRTs, the accessible parameter space is typically restricted to Regions I through VIII (using the numbering of Stix \cite{STIX92}).
In some regions of the CMA diagram, wave propagation exhibits highly anisotropic structures known as resonance cones \cite{fisherResonanceConesField1969, swan03}. These structures are associated with a critical angle $\theta_{\mathrm{c}}$ for which the refractive index diverges. Taking this singular limit in \cref{eqt-disp_rel_theta}, one obtains the resonance condition:
\begin{align}
	\tan^2\left(\theta_{\mathrm{c}}\right) &= \lim_{n\to\infty}\tan^2\theta = -\frac{P}{S}. \label{eqt:resonance_angle}
\end{align}
As the propagation angle approaches $\theta_{\mathrm{c}}$ along the resonant branch, progressively shorter wavelengths are supported by the plasma. As evidenced by \cref{eqt:resonance_angle}, a real resonance angle requires $P$ and $S$ to have opposite signs. This condition holds in CMA Regions III, VII, and VIII, which permit the formation of resonance cones. These regimes are commonly encountered in magnetized plasma devices, including ECR thrusters, making their numerical treatment directly relevant to practical applications.

The same resonance condition also has a mathematical interpretation. For a homogeneous collisionless plasma, the second-order vector wave equation yields a fourth-order scalar equation satisfied by each field component, as derived in \cref{app:principal_symbol}. Its principal symbol, obtained from the fourth-order spatial derivatives, is
\begin{align} \label{eqt:principal_eqt}
	\mathcal{Q}_4(\boldsymbol{\xi}) &= (\xi_x^2+\xi_z^2)(S\xi_x^2+P\xi_z^2), \qquad \bm{\xi}\in\mathbb{R}^2. 
\end{align}
The second factor also arises in the quasi-electrostatic limit of resonance-cone waves \cite{fisherResonanceConesField1969}.
When $P$ and $S$ have the same nonzero sign, the terms $S\xi_x^2$ and $P\xi_z^2$ cannot cancel. The principal symbol is therefore nonzero for every $\bm{\xi}\neq\bm{0}$, and the scalar equation is elliptic \cite{tayl91}.
In resonance-cone regimes, their opposite signs allow these terms to cancel along the directions defined by \cref{eqt:resonance_angle}, even at arbitrarily large wavenumbers. 
The scalar equation then loses ellipticity, a feature also discussed for reduced cold plasma models by Otway \cite{otwa10}, and the usual elliptic control of short spatial scales no longer applies. This can increase the sensitivity of the computed field to short-wavelength discretization errors and mesh geometry, motivating the comparison of numerical formulations and regularization strategies presented in this work.

\subsection{Fourier modal expansion on two-dimensional domains}

In the remainder of the present work, we are interested in characterizing wave-plasma simulations that are either 2D planar or 2D axisymmetric.
To study wave propagation in such 2D domains, it is convenient to model the fields in the out-of-plane direction with a modal expansion. 
Because the problem is linear, the modes can be solved independently and the full 3D field reconstructed efficiently by linear superposition, which is significantly faster than a direct 3D computation.
This out-of-plane expansion takes specific forms depending on the geometry of the domain.
For standard 2D planar, Cartesian problems in $(x,y,z)$ (such as the numerical results presented in \cref{SEC4_Fresnel}), the out-of-plane direction is $y$. We therefore expand the fields as
\begin{equation}
    \bm{\hat{E}}(x,y,z) = \int_{-\infty}^{\infty} \bm{E}^{(k_y)}(x,z) \exp(i k_y y) \:\mathrm{d} k_y,
\end{equation}
with $k_y$ the out-of-plane wavenumber.
The curl operator in this case reduces to
\begin{equation}
    \nabla \times  \rightarrow \left( \nabla_\mathrm{t} + ik_y\mathbf{\hat{e}}_y \right) \times, \qquad \text{with} \quad \nabla_\mathrm{t} \times \bm{E} = - \partial_z E_y \: \mathbf{\hat{e}}_x + \left(\partial_z E_x - \partial_x E_z \right) \mathbf{\hat{e}}_y + \partial_x E_y \: \mathbf{\hat{e}}_z,
\end{equation}
$\nabla_\mathrm{t}$ representing the in-plane part of the differential operator. 
This gives the curl-curl operator in the planar case:
\begin{equation}
    \nabla\times(\nabla\times\bm{E}) \rightarrow \nabla_\mathrm{t}\times(\nabla_\mathrm{t}\times\bm{E}) + ik_y\left[\nabla_\mathrm{t}\times(\mathbf{\hat{e}}_y\times\bm{E}) + \mathbf{\hat{e}}_y\times(\nabla_\mathrm{t}\times\bm{E})\right] - k_y^2\mathbf{\hat{e}}_y\times(\mathbf{\hat{e}}_y\times\bm{E}).
\end{equation}
On the other hand, for axisymmetric, cylindrical problems in $(r,\theta,z)$ (as presented in \cref{SEC5_ECRT}), the out-of-plane direction is the azimuthal angle $\theta$. In this case, the  fields can be expressed as 
\begin{equation}
    \bm{\hat{E}}(r,\theta,z) = \sum_{m=-\infty}^\infty \bm{E}^{(m)}(r,z) \exp(i m \theta)
\end{equation}
where $m$ is the azimuthal Fourier mode number.
The curl operator in this case becomes:
\begin{align}
    \nabla\times  \rightarrow \left[\nabla_\mathrm{t} + (im/r) \: \mathbf{\hat{e}}_\theta\right]\times, \qquad \text{with} \quad \nabla_\mathrm{t} \times \bm{E} = - \partial_z E_\theta \: \mathbf{\hat{e}}_r + \left(\partial_z E_r - \partial_r E_z \right) \mathbf{\hat{e}}_\theta + \left(E_\theta/r + \partial_r E_\theta \right) \mathbf{\hat{e}}_z.
\end{align}
The curl-curl operator in cylindrical coordinates is:
\begin{equation}
    \nabla\times(\nabla\times\bm{E}) \rightarrow \nabla_\mathrm{t}\times(\nabla_\mathrm{t}\times\bm{E}) + \frac{im}{r}\left[\nabla_\mathrm{t}\times\left(\mathbf{\hat{e}}_\theta\times\bm{E}\right) + \mathbf{\hat{e}}_\theta\times(\nabla_\mathrm{t}\times\bm{E})\right] - \frac{m^2}{r^2}\mathbf{\hat{e}}_\theta\times(\mathbf{\hat{e}}_\theta\times\bm{E})-\frac{im}{r^2}E_r\mathbf{\hat{e}}_\theta .
\end{equation}

% ------------------------------------------------------

% ------------------------------------------------------
\section{Numerical framework} \label{SEC3_discretization}
To investigate electromagnetic wave propagation in a magnetized plasma, we introduce a custom FEM solver for the frequency-domain wave equation \cref{eqt-curl_curl_gen}. This solver performs full-wave simulations of the electric field in 2D domains, utilizing the FEM ecosystem provided by the \texttt{Gridap.jl} library.
The assembled complex-valued sparse linear systems are solved by direct LU factorization using UMFPACK through \texttt{Gridap}'s default linear solver.
This implementation has been verified against the previous solver used in \cite{jime23a}, preserving the underlying physical model, while providing the modularity required to deploy the generalized anisotropic waveport boundary conditions described in \cref{subsec3_BC}. Furthermore, this transition to a more efficient computational framework yields substantial improvements in computational efficiency overall.

\subsection{Mixed-element discretization} \label{subsec_mixed_elements}

Nédélec edge elements \cite{nede80, nede86} form an $H(\mathrm{curl}, \Omega)$-conforming space by associating degrees of freedom (DoFs) with mesh edges rather than nodes, enforcing tangential continuity while permitting normal discontinuities.
This vector basis belongs to discrete de Rham sequences that preserve the relations between the differential operators at the discrete level, effectively mitigating spurious modes \cite{MONK03, boffi_mixed_2013, garc00}, which makes it well suited for the FE discretization of electromagnetic wave equations \cite{sun1995}.
In the context of the present 2D wave solver, the Fourier expansion of the fields in the out-of-plane coordinate leads naturally to a mixed-element discretization. The in-plane vector components, lying in the propagation plane, are discretized using the Nédélec space $V_h^1\subset H(\mathrm{curl},\Omega)$. By contrast, the out-of-plane component behaves effectively as a scalar within the Fourier expansion and can therefore be discretized using the Lagrange nodal space $V_h^0\subset H^1(\Omega)$.
This combination allows all three electric field components to be retained on a two-dimensional mesh within the Fourier-reduced formulation.

We implement this discretization by constructing a mixed functional space $V_h^1 \times V_h^0$ that maps both bases into a single global system.
When only a few Fourier modes are required, this dimensional reduction can substantially reduce the number of DoFs and the computational cost compared with a full three-dimensional Nédélec discretization.
For each Fourier mode, we define the Nédélec vector field $\bm{N}_h$ and the Lagrange scalar field $L_h$ as
\begin{align}
	\bm{N}_h &= \sum_i \bm{N}_i e_{\mathrm{t},i}, \qquad L_h = \sum_j L_j e_{\mathrm{out},j}.
\end{align}
Here, $e_{\mathrm{t},i}$ and $e_{\mathrm{out},j}$ are the unknown coefficients associated with the vector and scalar basis functions, respectively. The out-of-plane coefficients are $e_{\mathrm{out},j}=e_{y,j}$ in planar geometry and $e_{\mathrm{out},j}=e_{\theta,j}$ in cylindrical geometry. Modal superscripts are omitted from these coefficients and auxiliary fields for readability.
The Nédélec vector field is embedded in the computational plane, with $N_{1,h}$ along $z$ and $N_{2,h}$ along $x$ in planar geometry or $r$ in cylindrical geometry. For the planar configurations considered in \cref{SEC4_Fresnel}, the discrete electric field is therefore
\begin{align}
	\bm{E}_h^{(k_y)} &= \bm{N}_h + L_h\mathbf{\hat{e}}_y = N_{2,h}\mathbf{\hat{e}}_x + L_h\mathbf{\hat{e}}_y + N_{1,h}\mathbf{\hat{e}}_z. \label{eq:discrete_field_planar}
\end{align}
In cylindrical geometry, the reconstruction depends on the azimuthal mode number and must accommodate the regularity conditions at the symmetry axis. With the adopted dependence $\exp(im\theta)$, the modal field is represented as
\begin{align}
	\bm{E}_h^{(m)} &= \begin{cases} \bm{N}_h + L_h\mathbf{\hat{e}}_\theta = N_{2,h}\mathbf{\hat{e}}_r + L_h\mathbf{\hat{e}}_\theta + N_{1,h}\mathbf{\hat{e}}_z, & m=0,\\ 
    r\bm{N}_h + L_h(\mathbf{\hat{e}}_\theta \mp i\mathbf{\hat{e}}_r) = (rN_{2,h}\mp iL_h)\mathbf{\hat{e}}_r + L_h\mathbf{\hat{e}}_\theta + rN_{1,h}\mathbf{\hat{e}}_z, & m=\pm1,\\
    r\bm{N}_h + L_h\mathbf{\hat{e}}_\theta = rN_{2,h}\mathbf{\hat{e}}_r + L_h\mathbf{\hat{e}}_\theta + rN_{1,h}\mathbf{\hat{e}}_z, & |m|>1. \end{cases} \label{eq:discrete_field_cylindrical_modes}
\end{align}
The same reconstruction is applied to the trial and test functions. For the mode $m=+1$ considered in \cref{SEC5_ECRT}, it enforces $E_z|_{r=0}=0$ and $(E_r+iE_\theta)|_{r=0}=0$ by construction.
The anisotropic dielectric tensor $\bm{\kappa}$ couples the in-plane and out-of-plane electric field components, which are therefore solved together.

%%%%%%%%%%%%%%%%%%%%%%%%%%%%%%%%%%%%%%%%%%%%%%%%%%%%%%%%

\subsection{Standard wave formulation} \label{subsec:weak_formulation_W}

The FE framework relies on transforming the strong form \cref{eqt-curl_curl_gen} into a weak (variational) formulation, through the definition of a suitable functional space for the electric field.
The wave equation is multiplied by a complex-valued vector test function $\bm{\hat{T}}$ and integrated over the domain $\Omega$, leading to the sesquilinear form: 
\begin{equation} \label{eqt-weak_form_1}
\mathcal{A}\big(\bm{\hat{E}},\bm{\hat{T}}\big)=\int_\Omega \bm{\hat{T}^*} \cdot \left[\nabla \times \big(\nabla \times \bm{\hat{E}}\big)\right] \dV -k_0^2 \int_\Omega \bm{\hat{T}^*}  \cdot \bm{\kappa}\bm{\hat{E}} \dV,
\end{equation}
with $k_0 = \omega/c$ the vacuum wavenumber. The unknown trial electric field $\bm{\hat{E}}$ is approximated using a finite set of basis functions, while the test functions $\bm{\hat{T}}$ serve to weight the residual of the equation to zero in an integral sense. 
Using the same FE approximation spaces for the trial and test fields, with the appropriate boundary conditions, defines a conforming Galerkin method.
The primary differential term in \cref{eqt-weak_form_1} is integrated by parts, yielding a boundary contribution. The resulting weak problem is to find $\bm{\hat{E}}$, satisfying the imposed boundary conditions, such that, for every admissible test field $\bm{\hat{T}}$,
\begin{equation} \label{eqt-weak_form_2}
    \mathcal{A}_{\mathrm{W}}\big(\bm{\hat{E}},\bm{\hat{T}}\big)=\int_\Omega \left[\big(\nabla \times \bm{\hat{T}^*}\big)\cdot\big(\nabla \times \bm{\hat{E}}\big)-k_0^2 \bm{\hat{T}^*} \cdot \bm{\kappa}\bm{\hat{E}}\right] \dV + \oint_{\partial \Omega} \left[\mathbf{\hat{n}}\times\big(\curl\bm{\hat{E}}\big)\right]\cdot\bm{\hat{T}^*}\dS = 0,
\end{equation}
with $\mathbf{\hat{n}}$ the unit normal vector pointing outward from the boundary $\partial \Omega$.
Upon discretization, this infinite-dimensional weak form is restricted to a finite-dimensional FE space, transforming the integral equations into a global linear system where the unknowns are the discrete DoFs of the electric field. This formulation is denoted $\mathrm{W}$.

%%%%%%%%%%%%%%%%%%%%%%%%%%%%%%%%%%%%%%%%%%%%%%%%%%%%%%%%%%%%%%%%%%

\subsection{Saddle-point formulation} \label{subsec_Saddle_point}

We also consider a saddle-point formulation $\mathrm{SP}$, in which a scalar Lagrange multiplier $p$ imposes Gauss's law weakly \cite{demk98,grei07,jian14}, while retaining the electric-field approximation spaces of $\mathrm{W}$. For compatible spaces and boundary conditions, $\mathrm{SP}$ and $\mathrm{W}$ are equivalent in the sense that they yield the same discrete electric field whenever the discrete saddle-point system is uniquely solvable.
The additional multiplier nevertheless changes the size and block structure of the assembled system, potentially affecting the robustness and cost of its numerical solution. We therefore investigate whether this alternative algebraic structure offers numerical advantages across the plasma regimes considered, including those affected by numerical pollution.
These comparisons include collisionless cases, which lie outside the scope of the well-posedness analysis of Back et al. \cite{back15} for divergence-constrained formulations in magnetized plasmas, conducted under assumptions including strictly positive collisionality.

The corresponding strong form of the saddle-point formulation is
\begin{equation}
    \begin{cases}
        \nabla \times \big(\nabla \times \bm{\hat{E}}\big) - k_0^2 \bm{\kappa}\bm{\hat{E}} - \bm{\kappa}^{\dagger} \nabla p &= \bm{0}, \\
        \nabla \cdot \big(\bm{\kappa}\bm{\hat{E}}\big) &= 0.
    \end{cases}
\end{equation}
Here, $\bm{\kappa}^{\dagger}=(\bm{\kappa}^{*})^{\mathsf{T}}$ denotes the Hermitian adjoint of the dielectric tensor. This choice makes the constraint and multiplier blocks adjoints of one another, ensuring a self-adjoint off-diagonal coupling in the discrete saddle-point system.
In the weak framework, $\bm{\hat{E}}$ and $\bm{\hat{T}}$ denote the trial and test electromagnetic fields as previously, while $p$ and $q$ denote the trial and test Lagrange multipliers.
The weak problem is thus to find $(\bm{\hat{E}},p)$, satisfying the imposed boundary conditions, such that for every admissible test pair $(\bm{\hat{T}},q)$
\begin{equation}
\begin{aligned}
    \mathcal{A}_{\mathrm{SP}}\big((\bm{\hat{E}},p),(\bm{\hat{T}},q)\big) = &\int_{\Omega} \left[ \big(\nabla \times \bm{\hat{T}}^*\big) \cdot \big(\nabla \times \bm{\hat{E}}\big) - k_0^2 \bm{\hat{T}}^* \cdot \bm{\kappa} \bm{\hat{E}} - \bm{\hat{T}}^* \cdot \bm{\kappa}^{\dagger} \nabla p - \nabla q^* \cdot \bm{\kappa} \bm{\hat{E}} \right] \dV \\
    &+ \oint_{\partial \Omega} \left[\mathbf{\hat{n}} \times \big(\nabla \times \bm{\hat{E}}\big)\right] \cdot \bm{\hat{T}}^* \dS + \oint_{\partial \Omega} q^* \big(\mathbf{\hat{n}} \cdot \bm{\kappa} \bm{\hat{E}}\big) \dS = 0.
    \label{eqt:weak_form_SP}
\end{aligned}
\end{equation}
The complete discrete trial space comprises three functional components: the Nédélec edge space $V_h^1$ for the in-plane vector field, the nodal Lagrange space $V_h^0$ for the out-of-plane field, and an additional nodal space $Q_h^0\subset H_0^1(\Omega)$ for the multiplier, which uses the same Lagrange polynomial degree as the out-of-plane field.

%%%%%%%%%%%%%%%%%%%%%%%%%%%%%%%%%%%%%%%%%%%%%%%%%%%%%%%%%%%%%%%%%%

\subsection{Numerical configurations}
The mixed FE discretization is further specified by the mesh geometry and polynomial approximation order. We introduce a common notation to identify the numerical configurations considered throughout this work.
Triangular elements are implemented on unstructured meshes generated using a Frontal-Delaunay algorithm. This meshing strategy naturally yields a nearly hexagonal elemental arrangement in the bulk of the domain, for which favorable numerical dispersion properties have been reported in homogeneous wave-propagation problems \cite{mullen_dispersion_1982,warr95}, while retaining the geometric flexibility required to accurately represent the complex boundaries encountered in ECRT simulations. For comparison, rectangular elements are also considered later on structured Cartesian grids, yielding a regular and aligned discretization. 
For both geometries, finite elements of approximation order $\ell=1$ and $\ell=2$ are evaluated. The same nominal order $\ell$ is used for the Nédélec edge space and the Lagrange nodal spaces, simplifying the implementation and the comparison between discretizations.
The choice between $\ell=1$ and $\ell=2$ reflects the usual trade-off between computational cost and accuracy, with higher-order elements generally reducing numerical dispersion and phase errors (as shown in \cref{SEC4_Fresnel}) at increased computational complexity.

The numerical schemes considered here are identified by mesh geometry and approximation order. Triangular discretizations, used for most calculations, are written as $\mathcal{T}^\ell$, with $\ell\in\{1,2\}$, while rectangular elements are briefly introduced in \cref{subsec4_spectral_analysis} and denoted by $\mathcal{R}^\ell$. 
In \cref{subsec4_regionI}, where the standard wave formulation (\cref{subsec:weak_formulation_W}) and the saddle-point formulation (\cref{subsec_Saddle_point}) are compared explicitly, they are distinguished by the subscripts $\mathrm{W}$ and $\mathrm{SP}$, respectively. 
Thereafter, the formulation subscript is omitted and the notation $\mathcal{T}^\ell$ is used. 
When artificial damping is introduced, the corresponding numerical collisionality parameter $Z_{\mathrm n}$ is specified explicitly as an argument; otherwise $Z_{\mathrm n}=0$ is assumed, corresponding to no additional numerical damping. This is the collisionless case in the Fresnel benchmark (\cref{SEC4_Fresnel}), whereas physical collisions are retained in the ECRT application (\cref{SEC5_ECRT}).

%%%%%%%%%%%%%%%%%%%%%%%%%%%%%%%%%%%%%%%%%%%%%%%%%%%%%%%%%%%%%%%%%%

\subsection{Boundary conditions} \label{subsec3_BC}

Several boundary condition types are required to complete the variational formulations, in both the cylindrical (\cref{SEC5_ECRT}) and planar (\cref{SEC4_Fresnel}) configurations:
\paragraph{Perfect Electric Conductor (PEC) ---}
To model any metallic walls of the domain in the axisymmetric simulations, PEC boundary conditions are applied. This condition enforces the vanishing of the tangential component of the electric field on the boundary: $\mathbf{\hat{n}} \times \bm{\hat{E}} = \bm{0}$.
\paragraph{Axis symmetry ---}
Regularity at the symmetry axis $r=0$ is enforced through the modal field representation introduced in \cref{eq:discrete_field_cylindrical_modes}. For the mode $m=+1$ considered here, the trial and test spaces incorporate these conditions by construction, so no additional boundary constraint on the electric field is required at the axis.
\paragraph{Periodic boundaries ---}
In the planar Fresnel configuration, periodic boundary conditions are applied along the $x$-direction by pairing the upper and lower boundaries, in order to enforce the analytical periodicity of the solution in $x$. Within the FE discretization, this periodicity is handled natively by linking the DoFs of corresponding edges.
\paragraph{Lagrange multiplier Dirichlet condition ---}
For the saddle-point formulation, a homogeneous Dirichlet condition $p=0$ is imposed on the entire physical boundary $\partial\Omega$, such that the Lagrange multiplier and its test function belong to $H^1_0(\Omega)$. Consequently, the boundary integral arising from the divergence constraint in \cref{eqt:weak_form_SP} vanishes, since $q=0$ on $\partial\Omega$. For $k_0\neq0$, the source-free wave equation already implies Gauss's law, so that the continuous solution also satisfies the constrained system with $p=0$. At the discrete level, the multiplier then serves to explicitly enforce this constraint against the scalar test space.
\paragraph{Generalized waveport boundary conditions ---}
To model wave injection we use a custom waveport condition applicable to both isotropic and anisotropic media. It combines prescribed modal excitation with absorption of outgoing waves, thereby minimizing artificial reflections at the port.
This construction enables direct comparison with analytical solutions in planar configurations and extends naturally to the cylindrical geometry of the waveguide in ECRT simulations.
In classical electromagnetic problems the development of waveport boundary conditions for vacuum waveguides \cite{JIN09, pozar11} exploits modal orthogonality to separate the different wave components.

However, in anisotropic media such as a magnetized cold plasma, wave polarizations are generally non-orthogonal, and their modal coupling must be explicitly accounted for.
The following derivation proceeds in three steps: modal decomposition of the boundary field, projection onto the backward-propagating modes to extract reflection coefficients, and substitution into the boundary integral.
The boundary condition aims to represent the boundary integral term arising from the weak formulation \cref{eqt-weak_form_2} of the vector wave equation, along the waveport boundary $\Gamma\subset\partial\Omega$:
\begin{equation}
    I_{\Gamma} = \oint_{\Gamma} \left[\mathbf{\hat{n}}\times\big(\curl\bm{\hat{E}}\big)\right]\cdot\bm{\hat{T}^*}\dS.
\end{equation}
To simplify notation, we define the boundary operator $\bm{F}(\bm{\hat{E}}) = \mathbf{\hat{n}}\times(\curl\bm{\hat{E}})$, which corresponds (up to scaling constants) to the tangential magnetic trace of the field.
Assuming a modal decomposition at the waveport, the total electric field $\bm{\hat{E}}$ is expressed as
\begin{equation}
    \bm{\hat{E}} = \sum_{j=1}^{N} \left[a_j\bm{e}_j^{(+)} + b_j\bm{e}_j^{(-)}\right] = \bm{\Phi}^{(+)}\bm{a} + \bm{\Phi}^{(-)}\bm{b},
\end{equation}
where $\bm{a}=(a_1,\ldots,a_N)^{\mathsf{T}}$ and $\bm{b}=(b_1,\ldots,b_N)^{\mathsf{T}}$ contain the prescribed incident and unknown reflected amplitudes, respectively. The matrices $\bm{\Phi}^{(\pm)}$, of size $3 \times N$, collect the corresponding modal fields as columns:
\begin{equation}
    \bm{\Phi}^{(\pm)} = \begin{bmatrix} \bm{e}_1^{(\pm)} & \cdots & \bm{e}_N^{(\pm)} \end{bmatrix}.
\end{equation}
The analytical waveport modes depend on the local medium and geometry. For the generalized Fresnel problem in \cref{SEC4_Fresnel}, they are plane-wave solutions of the homogeneous anisotropic plasma dispersion relation, whereas for the ECRT simulations in \cref{SEC5_ECRT}, they are the standard vacuum circular-waveguide modes \cite{pozar11}. In both cases, these analytical solutions provide the modal fields $\bm{e}_j^{(\pm)}$.
By linearity, the boundary operator $\bm{F}$ follows an identical decomposition:
\begin{equation} \label{eqt_portBC_F}
    \bm{F}\big(\bm{\hat{E}}\big) = \bm{\Psi}^{(+)}\bm{a} + \bm{\Psi}^{(-)}\bm{b}, \qquad \textrm{with} \quad  \bm{\Psi}^{(\pm)} = \begin{bmatrix} \bm{F}\big(\bm{e}_1^{(\pm)}\big) & \cdots & \bm{F}\big(\bm{e}_N^{(\pm)}\big) \end{bmatrix}.
\end{equation}
To determine the reflected amplitudes $b_j$, we project the total field $\bm{\hat{E}}$ onto the backward-propagating modal set using the complex $L^2(\Gamma)$ inner product:
\begin{equation}
    \int_{\Gamma} \bm{\Phi}^{(-)^\dagger} \bm{\hat{E}}\dS = \left[\int_{\Gamma} \bm{\Phi}^{(-)^\dagger} \bm{\Phi}^{(+)}\dS\right]\bm{a} + \left[\int_{\Gamma} \bm{\Phi}^{(-)^\dagger} \bm{\Phi}^{(-)}\dS\right]\bm{b},
\end{equation}
where $\dagger$ denotes the conjugate transpose.
Let $\bm{V}(\bm{\hat{E}})$ be the vector of total-field projections, and let $\bm{A}$ and $\bm{B}$ be the modal coupling matrices, defined as:
\begin{equation}
    \bm{V}\big(\bm{\hat{E}}\big) = \int_{\Gamma} \bm{\Phi}^{(-)^\dagger}\bm{\hat{E}} \dS, \qquad \bm{A} = \int_{\Gamma} \bm{\Phi}^{(-)^\dagger}\bm{\Phi}^{(+)} \dS,  \qquad \bm{B} = \int_{\Gamma} \bm{\Phi}^{(-)^\dagger}\bm{\Phi}^{(-)} \dS.
\end{equation}
In isotropic-medium waveguides, modal orthogonality makes $\bm{B}$ diagonal; however, in the anisotropic case, the off-diagonal terms of $\bm{B}$ and $\bm{A}$ couple the different modes. 
The projection system reduces to $\bm{V}\big(\bm{\hat{E}}\big) = \bm{A}\bm{a} + \bm{B}\bm{b}$. Provided that the backward modal set is linearly independent on $\Gamma$, $\bm{B}$ is Hermitian positive definite and therefore invertible: 
\begin{equation}
    \bm{b} = \bm{B}^{-1}\left[\bm{V}\big(\bm{\hat{E}}\big) - \bm{A}\bm{a}\right].
\end{equation}
Substituting $\bm{b}$ into \cref{eqt_portBC_F} yields a closed expression for the boundary operator:
\begin{equation}
\bm{F}\big(\bm{\hat{E}}\big) = \bm{\Psi}^{(-)}\bm{B}^{-1}\bm{V}\big(\bm{\hat{E}}\big) + \left(\bm{\Psi}^{(+)}-\bm{\Psi}^{(-)}\bm{B}^{-1}\bm{A}\right)\bm{a}.
\end{equation}
This construction enables the injection of prescribed incident modes and the absorption of outgoing modes retained in the modal expansion, suppressing their reflection back into the domain, up to the accuracy of the FE discretization.
Finally, to incorporate this condition into the FE formulation, the boundary integral is decomposed into a sesquilinear form $P(\bm{\hat{E}}, \bm{\hat{T}})$ and a conjugate-linear source functional $U^{\mathrm{in}}(\bm{\hat{T}})$, following the nomenclature of classical FEM \cite{JIN09}:
\begin{equation}
    I_{\Gamma} = \oint_{\Gamma} \left[\mathbf{\hat{n}}\times\big(\curl\bm{\hat{E}}\big)\right]\cdot\bm{\hat{T}}^*\,\dS = P\big(\bm{\hat{E}},\bm{\hat{T}}\big) + U^{\mathrm{in}}\big(\bm{\hat{T}}\big).
\end{equation}
The trial-dependent term is
\begin{equation}
    P\big(\bm{\hat{E}},\bm{\hat{T}}\big) = \oint_{\Gamma} \left[\bm{\Psi}^{(-)}\bm{B}^{-1}\bm{V}\big(\bm{\hat{E}}\big)\right]\cdot\bm{\hat{T}}^*\,\dS,
\end{equation}
contributing to the global system matrix.
The source term responsible for wave injection is
\begin{equation}
    U^{\mathrm{in}}\big(\bm{\hat{T}}\big) = \oint_{\Gamma} \left[\left(\bm{\Psi}^{(+)}-\bm{\Psi}^{(-)}\bm{B}^{-1}\bm{A}\right)\bm{a}\right]\cdot\bm{\hat{T}}^*\,\dS.
\end{equation}

% ------------------------------------------------------

% ------------------------------------------------------
\section{Generalized Fresnel problem as a verification benchmark} \label{SEC4_Fresnel}
In this section, we develop the analytical solution to a generalized Fresnel problem as a verification benchmark for the wave solver and a reference for identifying and analyzing potential spurious numerical solutions. 
Whereas the classical Fresnel problem describes the interaction of an electromagnetic plane wave with an interface between two isotropic media, the anisotropic dielectric tensor of a magnetized cold plasma yields a dispersion relation with multiple branches, supporting two distinct transmitted eigenmodes that generally have different wavevectors.

\begin{figure}[pos=h!]
    \centering
    \begin{tikzpicture}[>=Stealth, font=\small]

    % --- Background ---
    % Plasma region
    \fill[plasmacol] (0, -2.5) rectangle (5.5, 2.5);

    % --- Interface ---
    % Dashed line at z=0
    \draw[thick, dashed] (0, -2.5) -- (0, 2.5);

    % --- Coordinate System ---
    % x and z axes
    \draw[->, very thick, >=stealth] (0,0.12) -- (0, 1) node[left] {$\hat{\mathbf{e}}_x$};
    \draw[->, very thick, >=stealth] (0.12,0) -- (1, 0) node[below right, xshift=-6pt] {$\hat{\mathbf{e}}_z$};

    % y axis (pointing out of the page)
    \draw[thick, fill=white] (0,0) circle (0.12);
    \fill (0,0) circle (0.04);
    \node[below left] {$\hat{\mathbf{e}}_y$};

    % --- Text Labels ---
    \node[anchor=east] at (-0.1, -2.3) {\textit{incident}};
    \node[anchor=east, text=bluevec] at (-0.1, 2.2) {\textit{reflected}};
    \node[anchor=east, text=redvec] at (5.4, -2.3) {\textit{transmitted}};

    % Material Boxes
    \node[draw, thick, inner sep=4pt, anchor=center, text height=1.6ex, text depth=0.25ex] at (-4.875, 2.215) {vacuum};
    \node[draw, thick, inner sep=4pt, anchor=center, text height=1.6ex, text depth=0.25ex] at (4.875, 2.215) {plasma};

    % --- Wave Vectors ---
    % Incident waves (1, 2) - Black
    \draw[->, very thick] (-3.6, -1.4) -- (-1.6, -0.2) node[midway, sloped, above, inner sep=2pt, font=\footnotesize, text opacity=0.65] {$\bm{k}_1^{(+)}$};
    \node[ inner sep=1pt, font=\normalsize] at (-3.85, -1.25) {$\bm{\tilde{E}}_1^{(+)}$};

    \draw[->, very thick] (-3.6, -2.) -- (-1.6, -0.8) node[midway, sloped, below, inner sep=2pt, font=\footnotesize, text opacity=0.65] {$\bm{k}_2^{(+)}=\bm{k}_1^{(+)}$};
    \node[ inner sep=1pt, font=\normalsize] at (-3.85, -2.35) {$\bm{\tilde{E}}_2^{(+)}$};

    % Reflected waves (3, 4) - Blue
    \draw[->, very thick, bluevec] (-1.6, 1.) -- (-3.6, 2.2)
    node[midway, sloped, above, inner sep=2pt, font=\footnotesize, text opacity=0.65, xshift=5pt] {$\bm{k}_1^{(-)}$};
    \node[bluevec, inner sep=1pt, font=\normalsize] at (-1.25, 1.2) {$\bm{\tilde{E}}_1^{(-)}$};
    
    \draw[->, very thick, bluevec] (-1.6, 0.4) -- (-3.6, 1.6)
    node[midway, sloped, below, inner sep=2pt, font=\footnotesize, text opacity=0.65, xshift=3pt] {$\bm{k}_2^{(-)}=\bm{k}_1^{(-)}$};
    \node[bluevec, inner sep=1pt, font=\normalsize] at (-1.25, 0.3) {$\bm{\tilde{E}}_2^{(-)}$};

    % Transmitted waves (5, 6) - Red
    \draw[->, very thick, redvec] (2.0, 0.2) -- (3., 1.4) node[midway, sloped, above, inner sep=3pt, font=\footnotesize, text opacity=0.65] {$\bm{k}_1^{(\mathrm{t})}$};
    \node[ redvec, inner sep=1pt, font=\normalsize] at (1.7, 0.25) {$\bm{\tilde{E}}_1^{(\mathrm{t})}$};

    \draw[->, very thick, redvec] (2.0, -1.0) -- (5.2, 0.2) node[midway, sloped, below, inner sep=3pt, font=\footnotesize, text opacity=0.65] {$\bm{k}_2^{(\mathrm{t})}$};
    \node[ redvec, inner sep=1pt, font=\normalsize] at (1.7, -0.98) {$\bm{\tilde{E}}_2^{(\mathrm{t})}$};

\end{tikzpicture}
    \caption{Schematic of a generalized Fresnel problem at the interface between vacuum and a uniform magnetized plasma. The incident electromagnetic wave in vacuum is decomposed into two orthogonal polarization components, with amplitudes $\bm{\tilde{E}}_1^{(+)}$ and $\bm{\tilde{E}}_2^{(+)}$, giving rise to reflected components with amplitudes $\bm{\tilde{E}}_1^{(-)}$ and $\bm{\tilde{E}}_2^{(-)}$. In the plasma, the anisotropic dielectric response supports two distinct transmitted eigenmodes, with amplitudes $\bm{\tilde{E}}_1^{(\mathrm{t})}$ and $\bm{\tilde{E}}_2^{(\mathrm{t})}$ and generally different wavevector directions; depending on the plasma regime, one of these modes may be evanescent.}
    \label{fig:4_Fresnel_setup}
\end{figure}
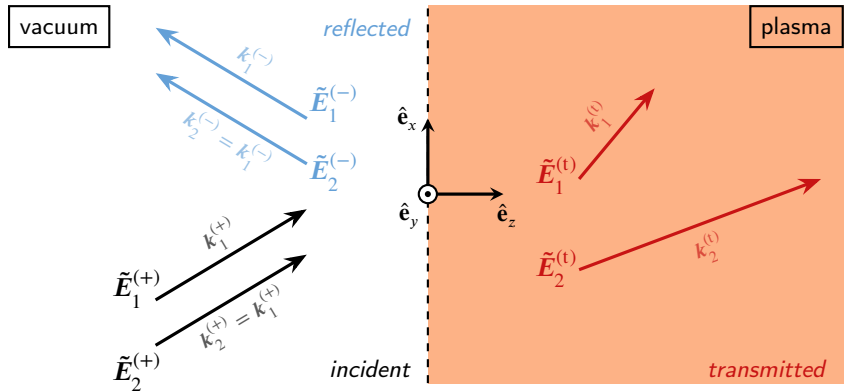

A schematic of the generalized Fresnel problem is shown in \cref{fig:4_Fresnel_setup}. The total electromagnetic field is decomposed into a superposition of six constituent plane waves: two incident and two reflected modes propagating within the vacuum region, and two transmitted eigenmodes propagating in the plasma. For concreteness, we fix $k_y=0$ for all the modes. This six-wave representation is used for the analytical developments presented below.

\subsection{Analytical solution} \label{subsec_4_ana}

We consider the planar electromagnetic scattering problem at the interface between two homogeneous, linear media: vacuum occupying the half-space $z<0$ (labeled $\mathrm{v}$), and a cold magnetized plasma occupying $z>0$ (labeled $\mathrm{p}$). The vacuum is characterized by an identity dielectric tensor $\id$, while the plasma is described by the frequency-dependent anisotropic dielectric tensor $\bm{\kappa}$, defined in \cref{eqt-dielectric_tensor}. The plasma is considered here in the collisionless (non-dissipative) regime. 
In both media, the magnetic permeability is assumed isotropic and equal to $\mu_0$, and the external magnetic field is assumed uniform and aligned with the $z$-axis --- as commonly adopted in cold plasma models --- although the formulation is not restricted to this orientation.
The interface is located at $z =0$, with the unit normal vector pointing into the plasma: $\mathbf{\hat{n}} = \mathbf{\hat{e}}_z$.

\paragraph{Plane wave representation and eigenmode structure ---}
The total electromagnetic field in each homogeneous region is expanded in terms of plane-wave eigenmodes. The electric fields of the incident, reflected, and transmitted modes are written as
\begin{equation} \label{eqt-general_Efield}
    \begin{aligned}
        \bm{E}_j^{(+)}(\bm{r},t) &= a_j\bm{e}_j^{(+)}(\bm{r})\exp(-i\omega t),\\
        \bm{E}_j^{(-)}(\bm{r},t) &= b_j\bm{e}_j^{(-)}(\bm{r})\exp(-i\omega t),\\
        \bm{E}_j^{(\mathrm{t})}(\bm{r},t) &= c_j\bm{e}_j^{(\mathrm{t})}(\bm{r})\exp(-i\omega t),
    \end{aligned}
\end{equation}
where $j=1,2$, the coefficients $a_j,b_j,c_j\in\mathbb{C}$ are the corresponding scalar amplitudes, and the modal fields are
\begin{equation}
    \bm{e}_j^{(s)}(\bm{r}) = \bm{P}_j^{(s)}\exp\left(i\bm{k}_j^{(s)}\cdot\bm{r}\right), \qquad s\in\{+,-,\mathrm{t}\}.
\end{equation}
Here, $\bm{P}_j^{(s)}$ denotes the normalized polarization vector. The associated constant electric-field amplitudes are therefore
\begin{equation}
    \bm{\tilde{E}}_j^{(+)} = a_j\bm{P}_j^{(+)}, \qquad \bm{\tilde{E}}_j^{(-)} = b_j\bm{P}_j^{(-)}, \qquad \bm{\tilde{E}}_j^{(\mathrm{t})} = c_j\bm{P}_j^{(\mathrm{t})}.
\end{equation}
For each medium, the admissible wavevectors $\bm{k}_j$ are obtained as solutions --- the allowed propagation roots --- of the continuous dispersion relation (\cref{eqt-disp_rel}) of the medium. For fixed angular frequency $\omega$ and transverse wavevector $(k_x,k_y)$, this relation defines a finite set of admissible roots for the normal component $k_z$, which may be real or complex, and thus include both propagating and evanescent modes.
For each $\bm{k}_j^{(s)}$, the associated polarization vectors $\bm{P}_j^{(s)}$ are defined as the nontrivial elements of the null space of the corresponding dispersion operator.
Translational invariance in the $(x,y)$-plane implies conservation of tangential wavevector components for all the six eigenmodes: $k_{j,x}^{(s)}=k_x$ and $k_{j,y}^{(s)}=k_y$. Therefore, at the interface ($z=0$), the common exponential phase factor $\exp[i(k_x x + k_y y - \omega t)]$ can be factored out entirely when enforcing boundary conditions. Consequently, the scattering problem reduces to relations between complex vector amplitudes $\bm{\tilde{E}}_j^{(s)}$ at the interface.

As established in \cref{fig:4_Fresnel_setup}, the total field is decomposed into six constituent plane waves: two incident and two reflected waves in the vacuum region, and two transmitted eigenmodes into the plasma. 
The general modal formulation admits two modes incident from the plasma half-space. In the present Fresnel problem, their amplitudes are set to zero; these modes are therefore omitted from the following development, and the plasma field contains only the two outgoing transmitted modes.
The total electric field amplitudes on the left (vacuum) and right (plasma) sides of the interface are 
\begin{equation}
    \bm{\tilde{E}}_{\mathrm{v}} = \sum_{j=1}^{2}\left(\bm{\tilde{E}}_j^{(+)}+\bm{\tilde{E}}_j^{(-)}\right), \qquad \bm{\tilde{E}}_{\mathrm{p}} = \sum_{j=1}^{2}\bm{\tilde{E}}_j^{(\mathrm{t})}.
\end{equation}
\paragraph{Interface conditions ---}

The solution is governed by the electromagnetic interface conditions derived from the integral forms of Maxwell's equations in the absence of free surface charges and surface currents. Although Maxwell’s interface conditions involve both tangential fields and normal fluxes, the latter are not independent once the plane-wave Maxwell equations are enforced. The scattering problem therefore reduces to the continuity of the tangential electric and magnetic fields across the interface:
\begin{subequations} \label{eqt:int_cond_tangential}
\begin{align}
    \mathbf{\hat{n}}\times\big(\bm{\tilde{E}}_{\mathrm{p}}-\bm{\tilde{E}}_{\mathrm{v}}\big) &= \bm{0}, \label{eqt:int_cond-E_strength}\\
    \mathbf{\hat{n}}\times\big(\bm{\tilde{H}}_{\mathrm{p}}-\bm{\tilde{H}}_{\mathrm{v}}\big) &= \bm{0}. \label{eqt:int_cond-H_strength}
\end{align}
\end{subequations}
To express the magnetic field condition entirely in terms of the electric field amplitudes, we apply Faraday's law of induction. For each time-harmonic plane wave, this directly yields
\begin{equation}
    \bm{\tilde{B}}_j^{(s)} = \frac{1}{\omega}\bm{k}_j^{(s)}\times\bm{\tilde{E}}_j^{(s)}.
\end{equation}
Substituting this relation alongside the assumption $\bm{B} = \mu_0 \bm{H}$ into the magnetic boundary condition (\cref{eqt:int_cond-H_strength}) yields
\begin{equation}
\mathbf{\hat{n}}\times\sum_{j=1}^{2}\left(\bm{k}_j^{(\mathrm{t})}\times\bm{\tilde{E}}_j^{(\mathrm{t})}\right) = \mathbf{\hat{n}}\times\sum_{j=1}^{2}\left(\bm{k}_j^{(+)}\times\bm{\tilde{E}}_j^{(+)}+\bm{k}_j^{(-)}\times\bm{\tilde{E}}_j^{(-)}\right).
\end{equation}

\paragraph{Linear system formulation ---}

Considering the two tangential components of \cref{eqt:int_cond-E_strength,eqt:int_cond-H_strength}, the four scalar interface conditions can be written compactly as 
\begin{subequations} \label{eqt:cond-tangential_all}
\begin{align}
\sum_{j=1}^{2}\left(\tilde{E}_{j,\alpha}^{(+)}+\tilde{E}_{j,\alpha}^{(-)}\right) &= \sum_{j=1}^{2}\tilde{E}_{j,\alpha}^{(\mathrm{t})}, \label{eqt:cond-Et}\\ \sum_{j=1}^{2}\left[\bm{k}_j^{(+)}\times\bm{\tilde{E}}_j^{(+)}+\bm{k}_j^{(-)}\times\bm{\tilde{E}}_j^{(-)}\right]_\alpha &= \sum_{j=1}^{2}\left(\bm{k}_j^{(\mathrm{t})}\times\bm{\tilde{E}}_j^{(\mathrm{t})}\right)_\alpha, \label{eqt:cond-Ht} 
\end{align}
\end{subequations}
where $\alpha\in \{x,y\}$.
Since $k_x$ and $k_y$ are common to both media, continuity of the tangential fields automatically ensures continuity of $\tilde{B}_z$ and $\tilde{D}_z$ through Faraday's and Ampère's laws. The normal interface conditions therefore impose no additional constraints.

Substituting the polarization expansions into \cref{eqt:cond-tangential_all} yields a $4\times4$ linear system for the reflected amplitudes $(b_1,b_2)$ and transmitted amplitudes $(c_1,c_2)$, with the incident amplitudes $(a_1,a_2)$ prescribed. When the system matrix is non-singular, its solution uniquely determines these amplitudes and hence the corresponding Fresnel coefficients.
Although the present formulation is specialized to a vacuum--cold-plasma interface, the derivation is directly extensible to two homogeneous anisotropic media under the same interface and radiation assumptions.

\subsubsection{Reference fields in selected plasma regimes}

\paragraph{Wave and plasma parameters --- } 
For all the configurations, the domain is excited by an incident plane wave from vacuum with modal amplitudes of $a_1 = a_2 = 1\,\mathrm{V/m}$ and a driving frequency of $\omega/2\pi = 5.8\,\mathrm{GHz}$. 
The wave propagates obliquely at a $45^\circ$ angle in the $zx$-plane, characterized by the normalized wavevector $\bm{n} = \bm{k}/k_0 = \sqrt{2}/2 (1, \, 0, \, 1)^\textrm{T}$. 
To excite the relevant plasma modes, we use two incident polarization states: a purely out-of-plane transverse electric (TE) wave with polarization vector $\bm{P}_1^{(+)} = (0, -1, 0)^\textrm{T}$, and an in-plane transverse magnetic (TM) wave with $\bm{P}_2^{(+)} = \sqrt{2}/2 (1, 0, -1)^\textrm{T}$.
The normalized parameters $X = \omega_{\mathrm{pe}}^2/\omega^2$ and $Y = \omega_{\mathrm{ce}}/\omega$ define the wave propagation regimes in plasma. The reference values for each of the CMA diagram regions studied here are summarized in \cref{tab:CMA_regions}.
\begin{table}[pos=h!]
\centering
\begin{tabular}{l@{\hspace{1cm}}cc}
\toprule
CMA Region & $X = \omega_{\mathrm{pe}}^2/\omega^2$ & $Y = \omega_{\mathrm{ce}}/\omega$ \\
\midrule
I    & 0.3 & 0.3 \\
VII  & 2.0 & 4.0 \\
VIII & 8.0 & 4.0 \\
\bottomrule
\end{tabular}
\caption{Dimensionless cold plasma parameters defining the specific CMA regions selected for the generalized Fresnel problem.}
\label{tab:CMA_regions}
\end{table}

Although the underlying physical model is collisionless, a spatially uniform artificial collision frequency can be introduced in the generalized Fresnel problem to regularize the numerical solution. This contribution is defined as $\nu = Z_{\mathrm n}\omega$, where $Z_{\mathrm n}$ is a dimensionless numerical collisionality parameter. Since physical collisions are absent, the total normalized collisionality in this section is $Z=\nu/\omega=Z_{\mathrm n}$.

\paragraph{Analytical field distributions --- }

\begin{figure}[pos=h!]
    \centering
    \includegraphics[width=0.8\linewidth]{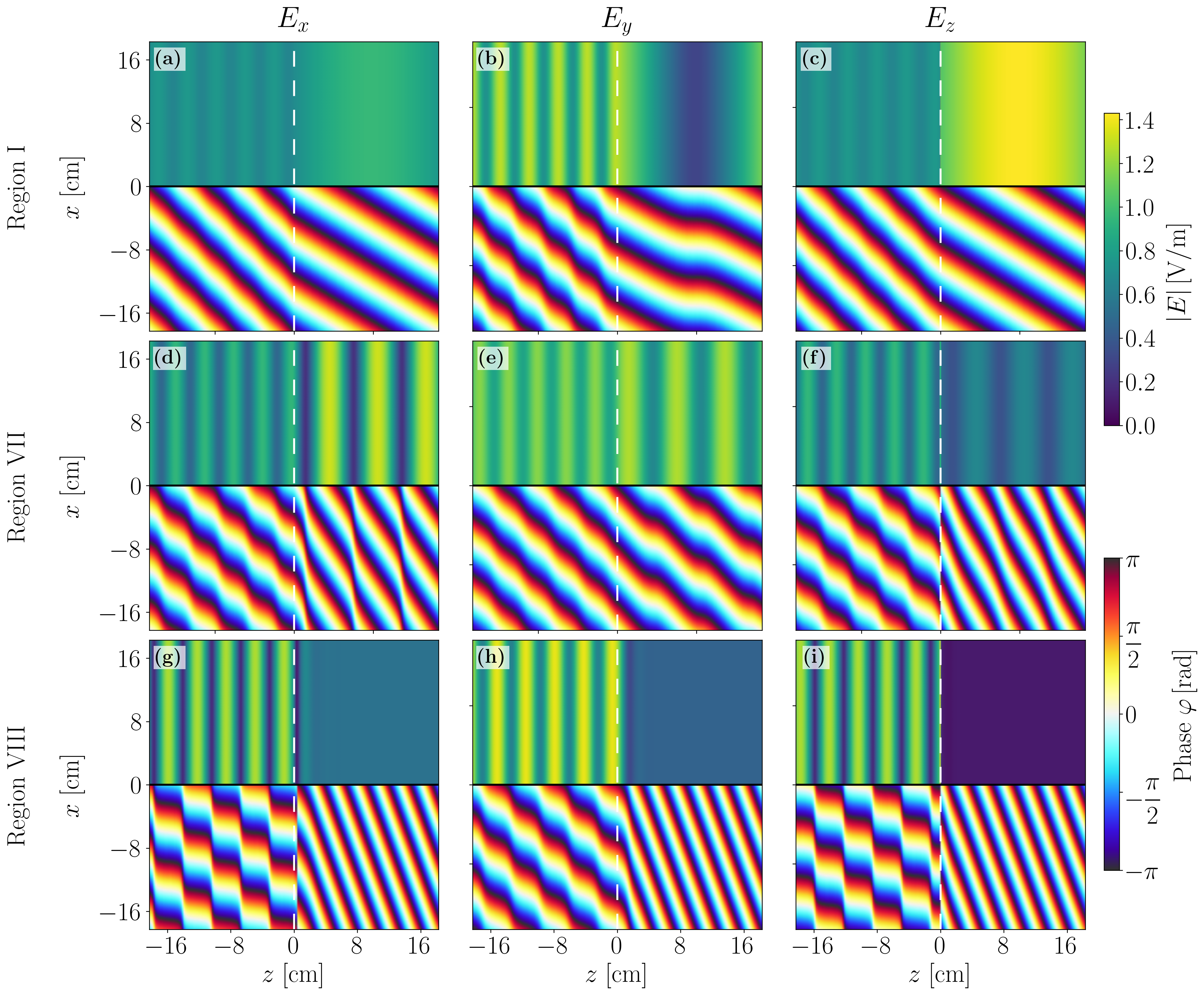}
    \caption{Analytical spatial distributions of the electric field components ($E_x, E_y, E_z$) for the generalized Fresnel problem, evaluated on the computational domain. Each panel is horizontally split: the top half displays the field magnitude $|E|$ (with common color scale set by Region I), and the bottom half displays the spatial phase $\varphi$. The vertical white dashed line at $z=0$ denotes the vacuum--plasma interface. Rows correspond to distinct regions from the CMA diagram.}
    \label{FIG4_analytical_ref}
\end{figure}

Using the wave and plasma parameters specified above, we evaluate the analytical solution to obtain reference fields for the subsequent numerical comparisons. \cref{FIG4_analytical_ref} presents the exact analytical spatial distributions of the electric field components ($E_x$, $E_y$, $E_z$) across the computational domain.

The main observation concerns the field structure in the resonance-cone regimes (Regions VII and VIII). For the selected parameters and excitation, the analytical fields remain smooth within each homogeneous region and exhibit no mesh-scale oscillatory structures.

\subsection{Numerical results} \label{subsec_4_results}

This section presents the numerical FEM results for the generalized Fresnel problem. Although reducible to one dimension, the problem is solved on a two-dimensional mesh to verify the planar FEM implementation and assess its sensitivity to mesh geometry and resolution against the analytical solution.
All simulations were performed on a cluster node equipped with two Intel Xeon Silver 4316 processors (40 physical cores in total, or equivalently 80 hardware threads) and $256\,\mathrm{GB}$ of RAM. The sparse linear systems were solved directly with UMFPACK through \texttt{Gridap}, with BLAS configured to use 10 threads during the solution stage; matrix assembly remained serial.

\paragraph{Mesh parameters --- } 

The analysis focuses on triangular meshes, particularly relevant for the complex geometries encountered in realistic wave-plasma simulations of EPTs. These meshes have an approximately uniform element size, with no local refinement tailored to the problem. The vacuum--plasma interface is explicitly represented by mesh edges and nodes, dividing the domain into two equal rectangular subdomains, with no elements crossing it. The influence of the element geometry is examined separately in \cref{subsec4_spectral_analysis}, where the corresponding results obtained with rectangular elements are presented.
The computational domain is a square in the $(z,x)$ plane, with periodic boundary conditions on the upper and lower edges. Exact periodicity of the analytical field requires the domain height $L_x$ to satisfy $k_xL_x=2\pi N$, where $N$ is an integer. The domain size is then directly dependent on the properties of the incident waves. Here, $L_x = L_z \approx 36.55\,\mathrm{cm}$, $N = 5$.
On the left and right boundaries, the waveport conditions developed previously in \cref{subsec3_BC} prescribe the incident modal amplitudes and allow the outgoing (transmitted and reflected) modes to leave the computational domain. These boundary conditions reproduce the analytical configuration of a wave incident on the interface between vacuum and a semi-infinite plasma.
Both first- ($\ell=1$) and second-order ($\ell=2$) mixed finite elements are evaluated, as defined in \cref{SEC3_discretization}.
The two reference triangular meshes are denoted by $M_{6\mathrm{k}}$ and $M_{15\mathrm{k}}$ --- referring to their number of elements --- with their main properties summarized in \cref{tab:mesh_params}.

\begin{table}[pos=h!]
\centering
\begin{tabular}{llcc}
\toprule
Parameter & Symbol & $M_{6\mathrm{k}}$ & $M_{15\mathrm{k}}$ \\
\midrule
Number of elements            & $N_h$               & 5,850  & 15,356 \\
Characteristic size           & $\Delta$            & 0.883   & 0.544   \\
Degrees of freedom ($\ell=1$) & $N_{\mathrm{DoFs}}$ & 11,800  & 30,874 \\
Degrees of freedom ($\ell=2$) & $N_{\mathrm{DoFs}}$ & 41,150 & 107,816 \\
\bottomrule
\end{tabular}
\caption{Spatial discretization parameters of the reference triangular meshes $M_{6\mathrm{k}}$ and $M_{15\mathrm{k}}$ for the Fresnel problem: number of elements, normalized characteristic mesh size $\Delta = k_0 h$, and global degrees of freedom for $\mathcal{T}_\mathrm{W}^1$ and $\mathcal{T}_\mathrm{W}^2$.}
\label{tab:mesh_params}
\end{table}

\subsubsection{Elliptic regime: CMA Region I} \label{subsec4_regionI}

\begin{figure}[pos=h!]
    \centering
    \includegraphics[width=0.8\linewidth]{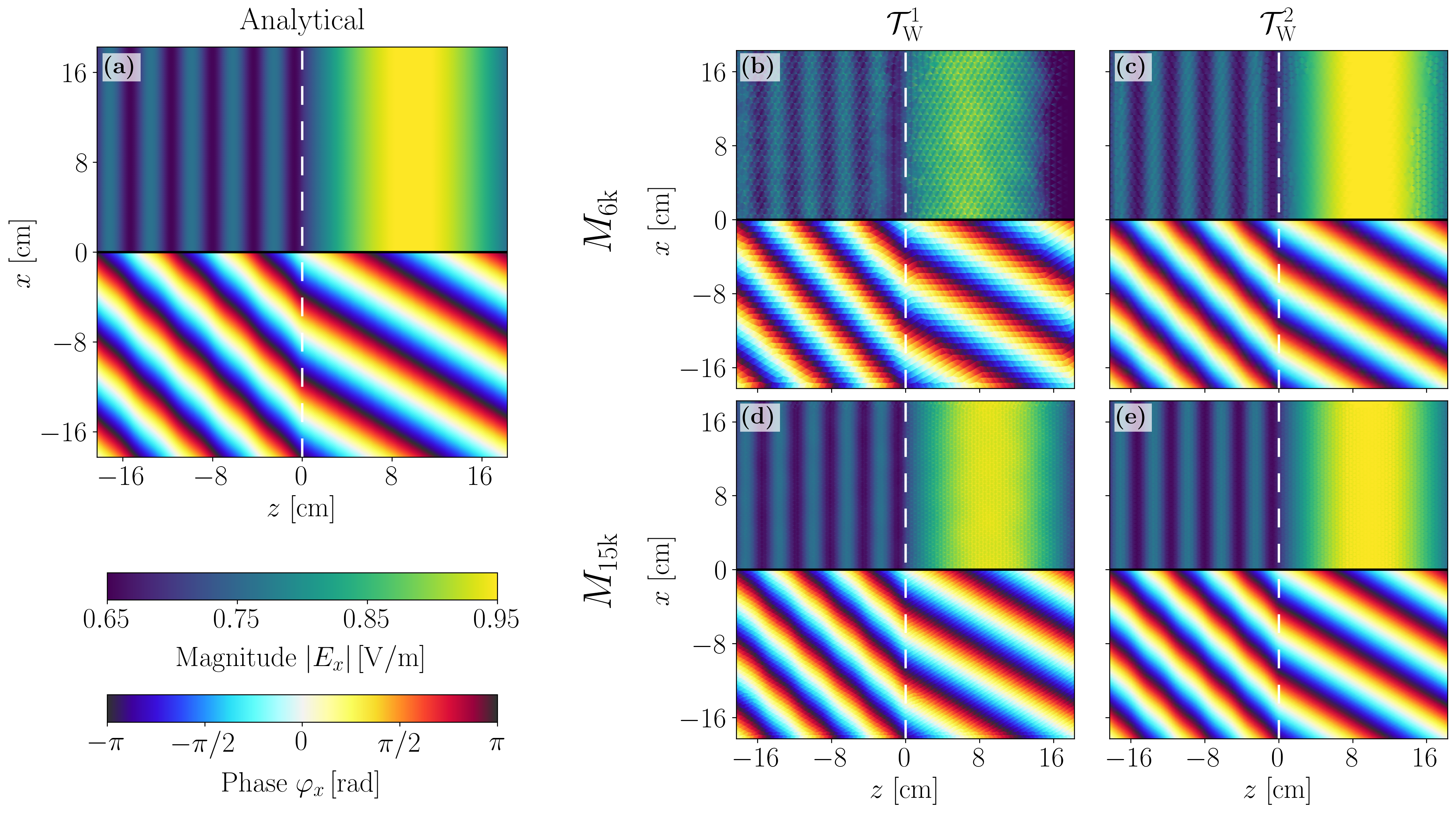}
    \caption{Spatial distributions of the $E_x$ component in CMA Region I. Panel (a) shows the analytical reference from \cref{FIG4_analytical_ref}(a). The remaining panels show the standard wave formulation with first-order elements $\mathcal{T}_\mathrm{W}^1$ [panels (b),(d)] and second-order elements $\mathcal{T}_\mathrm{W}^2$ [panels (c),(e)], computed on $M_{6\mathrm{k}}$ (top row) and $M_{15\mathrm{k}}$ (bottom row). Within each panel, the upper and lower halves display the magnitude $|E_x|$ and phase $\varphi_x$, respectively. The color scales are defined from the analytical solution and applied to the numerical results to facilitate direct comparison. Second-order elements closely reproduce the analytical wave structure at both resolutions, while the first-order results improve under mesh refinement.}
    \label{FIG4_R1_2Dmaps}
\end{figure}

We first assess the numerical formulations in CMA Region I, using the baseline plasma parameters detailed in \cref{tab:CMA_regions} and the mesh parameters described in \cref{tab:mesh_params}. Characterized by low plasma density, Region I does not feature resonance cones and provides an initial verification of the discretizations. The spatial distributions of the $E_x$ component of the electric field are presented in \cref{FIG4_R1_2Dmaps}, with the analytical solution in panel (a) as reference. Only the standard wave formulation results are shown, as the corresponding saddle-point maps are indistinguishable.

On the coarse $M_{6\mathrm{k}}$ mesh, the first-order discretization $\mathcal{T}_\mathrm{W}^1$ [panel (b)] exhibits visible discrepancies relative to the analytical solution, particularly in the magnitude distribution within the plasma. Refinement to $M_{15\mathrm{k}}$ [panel (d)] substantially improves the solution, although the peak magnitude within the plasma appears slightly lower than in the analytical reference.
Increasing the approximation order to $\ell = 2$ provides a substantial accuracy improvement at both resolutions [panels (c),(e)]. Our numerical tests indicate that this improvement stems primarily from the second-order Nédélec edge functions, as retaining first-order Lagrange elements does not significantly degrade the solution. 
To assess accuracy and computational cost, a mesh refinement convergence study is performed for each scheme and approximation order. Mesh convergence is evaluated using hierarchical nested refinement so that successive meshes differ only by element subdivision, minimizing changes in mesh topology and element quality.
The global accuracy of the numerical solutions is measured using the relative error in $H(\mathrm{curl})$, the natural functional space for our mixed elements formulation \cite{hipt02, zhon09}. This norm accounts for errors in both the electric field and its curl, and is defined as:
\begin{equation}
    \| \bm{E} \|_{H(\text{curl})} = \left( \int_{\Omega} \left( |\bm{E}|^2 + k_0^{-2}|\nabla \times \bm{E}|^2 \right) \mathrm{d}\Omega \right)^{1/2}
\end{equation}
over the computational domain $\Omega$. The factor $k_0^{-2}$ ensures that both contributions have the same physical units, using the vacuum wavenumber as the reference scale for the curl term.
The relative error $\epsilon_h$ is then given by:
\begin{equation}
    \epsilon_h = \frac{\| \bm{E}_h - \bm{E}_{\mathrm{ana}} \|_{H(\mathrm{curl})}}{\| \bm{E}_{\mathrm{ana}} \|_{H(\mathrm{curl})}}
\end{equation}
where $\bm{E}_h$ is the numerical solution of the electric field and $\bm{E}_{\mathrm{ana}}$ the analytical solution derived previously. 

\begin{figure}[pos=h!]
    \centering
    \includegraphics[width=0.9\linewidth]{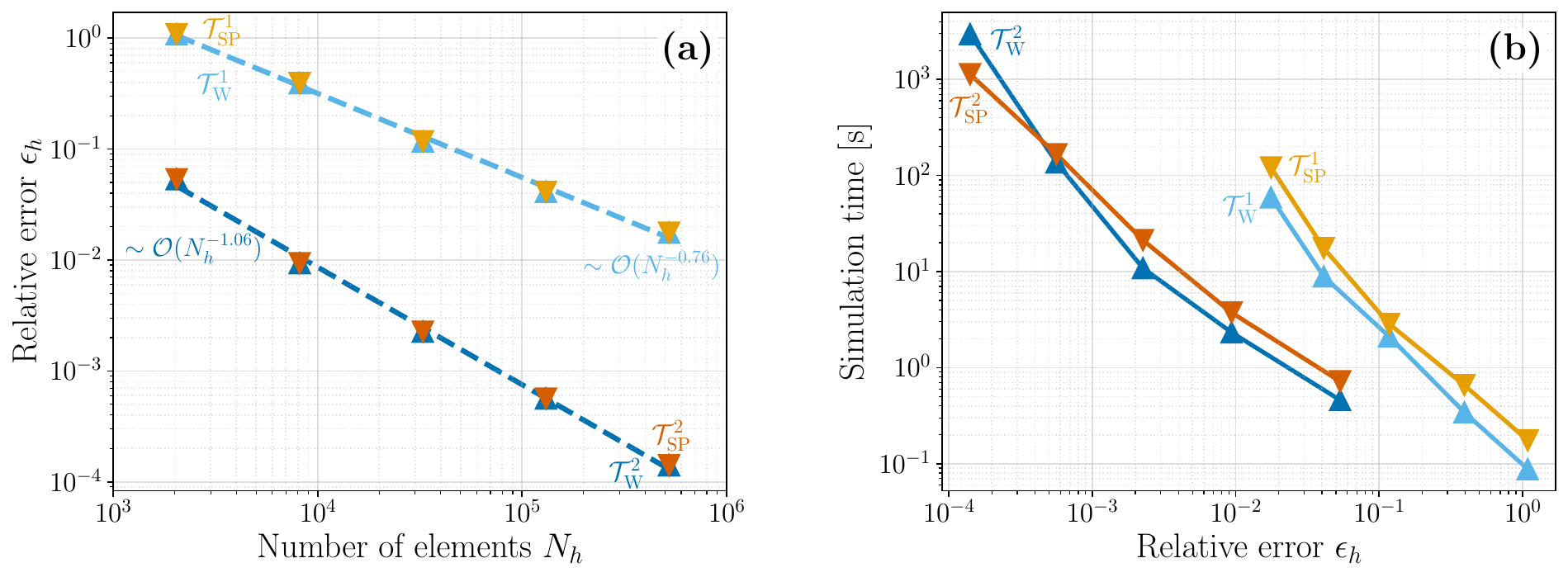}
    \caption{Convergence and computational performance of the standard wave $\mathcal{T}_\mathrm{W}$ and saddle-point $\mathcal{T}_\mathrm{SP}$ formulations with first- and second-order triangular discretizations in CMA Region I. Panel (a) shows the relative error $\epsilon_h$, measured in the $H(\mathrm{curl})$ norm, against the number of elements $N_h$. Dashed lines indicate power-law fits of the standard wave results, as the saddle-point results overlap at each order. Panel (b) shows the simulation time against the achieved relative error. Second-order discretizations provide the most favorable accuracy-cost trade-off over the tested range.}
    \label{FIG4_R1_convergence}
\end{figure}

\cref{FIG4_R1_convergence} presents the convergence analysis in Region I. Panel (a) shows the relative error $\epsilon_h$ defined above, as a function of the number of mesh elements $N_h$. Both first- and second-order discretizations exhibit steady error reduction under mesh refinement, with overlapping convergence curves for the standard wave-equation $\mathcal{T}_\mathrm{W}$ and saddle-point $\mathcal{T}_\mathrm{SP}$ formulations as expected due to their mathematical equivalence. 
This agreement is also observed in the other CMA regions tested, including resonance-cone regimes. At a given mesh resolution, second-order elements achieve substantially lower errors, showing the improvement in accuracy obtained by increasing the approximation order. 

For a quasi-uniform two-dimensional mesh, an error of order $\mathcal{O}(h^\ell)$ corresponds to $\mathcal{O}(N_h^{-\ell/2})$, since $N_h\propto h^{-2}$. The fitted second-order slope is approximately $-1.06$, close to the expected slope of $-1$ associated with second-order convergence in $h$. The first-order results yield an empirical slope of approximately $-0.76$, steeper than the nominal first-order rate ($-0.5$) over the sampled resolution range. 
The corresponding accuracy-versus-cost comparison is presented in \cref{FIG4_R1_convergence}(b), which plots the computational time against the achieved relative error. Here, computational time comprises matrix assembly and linear solution, excluding Julia's JIT compilation time. At a fixed approximation order, the assembly cost scales approximately linearly with the number of DoFs, whereas the linear solver time grows more rapidly under mesh refinement. Consequently, the linear solution accounts for most of the total computational time on the finest meshes.
For any comparable error levels within the tested range, the second-order discretizations require substantially less time than the first-order alternatives. 
Although the higher-order basis increases the cost per element, its accuracy allows considerably coarser meshes to be used. This behavior is consistent with the reduced dispersion error obtained by increasing the approximation order of Nédélec elements \cite{ainsworth2004}. Second-order discretization therefore provides the most favorable accuracy-cost balance among the tested configurations in Region I.

The choice of variational formulation (W or SP) does affect computational cost in our present implementation. On coarser meshes, the standard wave formulation solves faster, as expected from the additional DoFs introduced by the Lagrange multiplier in the saddle-point formulation.
For the second-order discretizations, however, a crossover occurs under mesh refinement: $\mathcal{T}_\mathrm{W}^2$ exhibits less favorable computational scaling and becomes slower than $\mathcal{T}_\mathrm{SP}^2$ at the finest resolutions considered.
This degradation is associated with fill-in during sparse LU factorization, performed here through \texttt{Gridap}'s default direct solver. Although the standard wave-equation formulation produces a smaller assembled matrix, the number of nonzero entries in its LU factors grows more rapidly with refinement than for the saddle-point formulation. The resulting fill-in increases both memory requirements and computational time, whereas the saddle-point formulation retains lower fill-in despite its additional unknowns. This comparison depends on the ordering and pivoting strategies employed and may vary across solvers.

Given their equivalence, the two formulations can therefore be used interchangeably from the standpoint of field accuracy, and the choice between them is only guided by computational efficiency: $\mathcal{T}_\mathrm{W}^2$ is preferable on coarser meshes, whereas $\mathcal{T}_\mathrm{SP}^2$ becomes advantageous on finer meshes and can enable convergence studies at resolutions where the standard formulation is limited by computational cost or memory requirements. 
Accordingly, the following sections use $\mathcal{T}^1$ and $\mathcal{T}^2$ to denote the first- and second-order triangular discretizations, without distinguishing between the two variational formulations.

%----------------------------------------------------------------------
\subsubsection{Resonance-cone regime: CMA Region VIII} \label{subsec_4_res_R8}
%----------------------------------------------------------------------

\begin{figure}[pos=h!]
    \centering
    \includegraphics[width=0.8\linewidth]{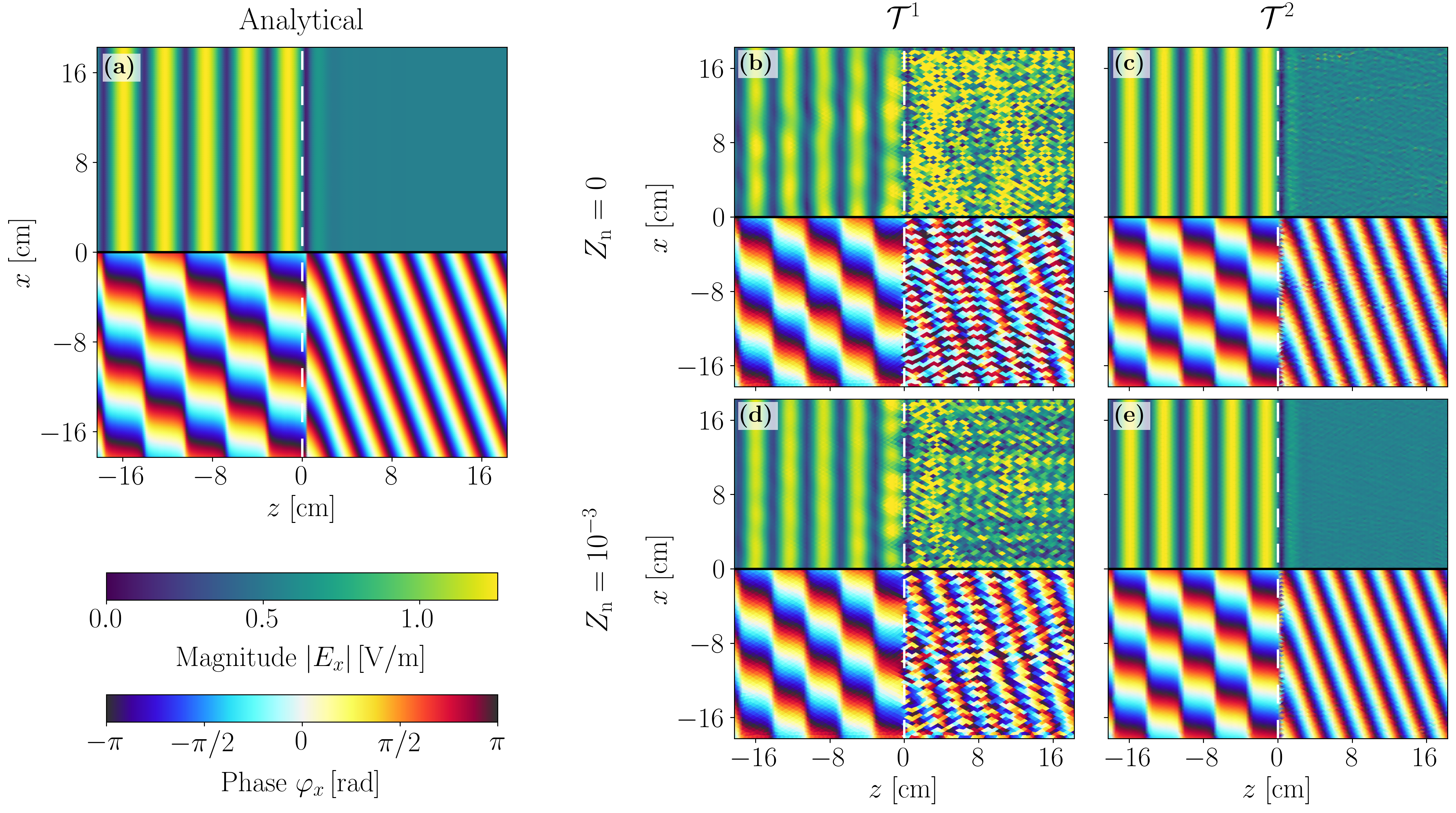}
    \caption{Spatial distributions of the $E_x$ component in CMA Region VIII. Panel (a) shows the collisionless analytical reference from \cref{FIG4_analytical_ref}(g). The remaining panels show the first-order triangular discretization $\mathcal{T}^1$ [panels (b),(d)] and the second-order discretization $\mathcal{T}^2$ [panels (c),(e)], all computed on the $M_{\text{15k}}$ mesh. The top row shows the collisionless case, $Z_{\mathrm n}=0$, whereas the bottom row includes additional numerical collisionality with $Z_{\mathrm n}=10^{-3}$. Within each panel, the upper and lower halves display the magnitude $|E_x|$ and phase $\varphi_x$, respectively. The color scales are defined from the analytical solution and applied unchanged to the numerical results to facilitate direct comparison.}
    \label{FIG4_R8_2Dmaps}
\end{figure}

We next examine CMA Region VIII, to investigate the numerical difficulties associated with resonance-cone regimes. Using the corresponding plasma parameters in \cref{tab:CMA_regions}, we compare the computed fields with the analytical solution and assess the effect of additional numerical collisionality.
The spatial distributions of the $E_x$ component are presented in \cref{FIG4_R8_2Dmaps}. The numerical results are shown on the $M_{15\mathrm{k}}$ mesh. 

Panel (a) shows the collisionless analytical reference, while the upper and lower rows of numerical results correspond to the collisionless $Z_{\mathrm n}=0$ [panels (b),(c)] and fixed collisionality $Z_{\mathrm n}=10^{-3}$ [panels (d),(e)] cases, respectively.
In the collisionless case, the first-order discretization $\mathcal{T}^1$ fails to reproduce the analytical field structure within the plasma. Both magnitude and phase are dominated by mesh-scale oscillations [\cref{FIG4_R8_2Dmaps}(b)], unlike the fields obtained in Region I. The second-order discretization $\mathcal{T}^2$ recovers the large-scale analytical wave pattern considerably more accurately [panel (c)], but retains weak short-wavelength oscillations.
These structures are absent from the analytical field for the same configuration, suggesting that the spurious oscillations encountered in cold plasma FE simulations are numerical artifacts rather than physical phenomena.
A small uniform numerical collisionality can mitigate this pollution. In the generalized Fresnel problem, this additional damping is introduced into the dielectric tensor through $\nu=Z_{\mathrm n}\omega$. For $Z_{\mathrm n}=10^{-3}$, the second-order solution [\cref{FIG4_R8_2Dmaps}(e)] exhibits a further reduction in mesh-scale spurious oscillations and closely reproduces the analytical wave structure. The first-order field remains strongly polluted at this resolution [panel (d)].

\begin{figure}[pos=h!]
    \centering
    \includegraphics[width=0.5\linewidth]{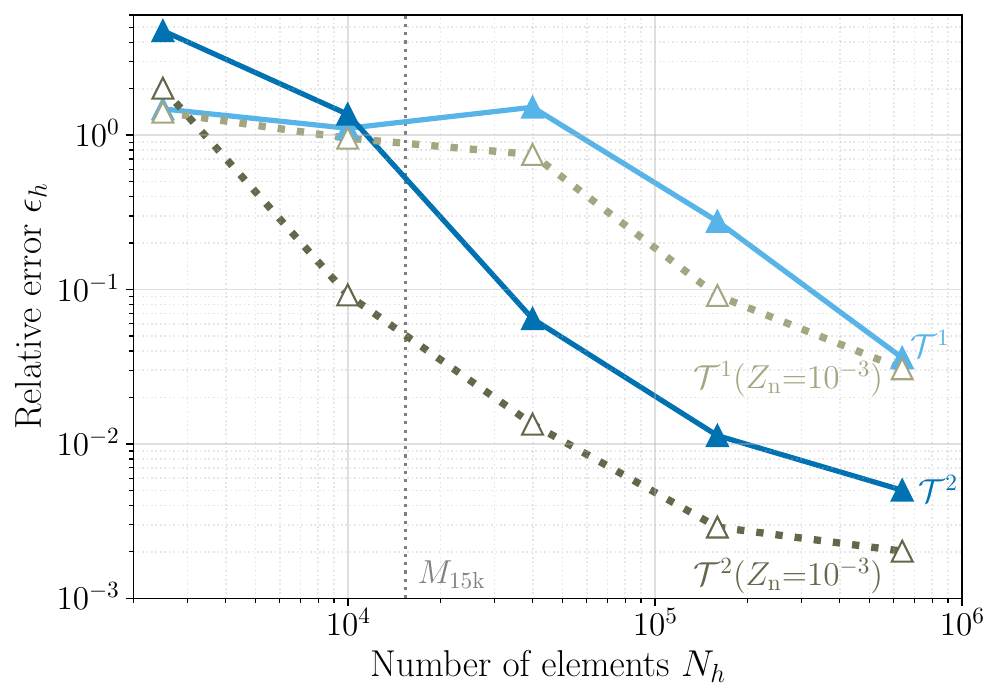}
    \caption{Convergence analysis of first- and second-order triangular discretizations in CMA Region VIII. The relative error $\epsilon_h$, measured against the collisionless analytical solution in the $H(\mathrm{curl})$ norm, is plotted against the number of elements $N_h$. Solid curves show the collisionless $\mathcal{T}^1$ and $\mathcal{T}^2$ results, while dotted curves include numerical collisionality $Z_{\mathrm n}=10^{-3}$. The coarsest points are retained for completeness, with consistent error reduction emerging on finer meshes. A vertical dotted line marks the reference $M_{15\mathrm{k}}$ resolution used in \cref{FIG4_R8_2Dmaps}(b)--(e). Added collisionality reduces the error, particularly for second-order elements, although the regularized curve appears to approach an asymptotic plateau at the finest resolutions.}
    \label{FIG4_R8_conv_1}
\end{figure}

To quantify these observations, \cref{FIG4_R8_conv_1} presents the relative error against the collisionless analytical solution as a function of $N_h$. The coarsest meshes give large errors, and asymptotic error reduction can be observed at finer resolutions. 
Although refinement reduces the first-order error, this does not establish that the spurious oscillations have disappeared. The global error provides a quantitative measure of accuracy, but should be interpreted together with the field maps when assessing the persistence of mesh-scale pollution.

For $\mathcal{T}^1$, additional collisionality lowers the error, but the improvement depends markedly on the mesh size. At the $M_{15\mathrm{k}}$ resolution marked by the vertical dotted line, the errors remain similarly large with and without damping, consistent with the observation in \cref{FIG4_R8_2Dmaps}.
At sufficiently fine resolutions, second-order elements achieve substantially lower errors than their first-order counterparts, and collisionality produces a further clear improvement (at least up to a limit value in $Z_{\mathrm{n}}$). However, both second-order curves show slower error reduction on the finest meshes.

In the collisionless case, this apparent error saturation is consistent with residual spurious oscillations limiting accuracy, while the regularized error also includes damping-induced changes that refinement does not remove at fixed collisionality, or at a rate slower than the expected convergence order.
In the collisional case, this can reflect the numerical solution approaching the corresponding collisional analytical solution, rather than the collisionless one used in the definition of the relative error.
The benefit of the added collisionality seems to vary with resolution: the collisionless and regularized curves are relatively close on coarse meshes, separate at intermediate resolutions, and approach one another again on the finest meshes, particularly for $\mathcal{T}^1$. This behavior shows that the same damping level provides different accuracy gains depending on mesh resolution and may become excessive under refinement, motivating the parametric study in \cref{subsec_4_res_coll}.

%----------------------------------------------------------------------
\subsection{Spectral diagnostics and influence of mesh geometry} \label{subsec4_spectral_analysis}
%----------------------------------------------------------------------

To understand the origin of the spurious oscillations in resonance-cone regimes, such as Region VIII, we return to the collisionless problem and focus on the second-order standard wave formulation $\mathcal{T}^2$ by combining plane-wave residual analysis with a Fast Fourier Transform (FFT) of the computed electric field.
We examine Regions I, III, VII, and VIII of the CMA diagram to compare the spectral behavior in regimes with and without resonance cones. The plasma parameters are those given in \cref{tab:CMA_regions}; for Region III, we use $X=\omega_{\mathrm{pe}}^2/\omega^2=0.9$ and $Y=\omega_{\mathrm{ce}}/\omega=0.9$.

\subsubsection{Plane-wave residual and Fourier analyses}

The analysis described in Appendix \ref{subsec3_NUM_DISP} provides insight into the effect of the computational mesh on wave propagation by evaluating the residual obtained when the assembled FE operator is applied to an interpolated continuous plane-wave. At fixed angular frequency $\omega$, the analysis explores the two-dimensional dimensionless wavevector space $(k_z h,k_x h)$, where $h$ is the effective mesh size.
The smallest singular value $\sigma_{\min}$ of the reduced residual matrix is computed throughout this parameter space. Its minima identify wavevectors for which the adopted plane-wave ansatz most closely satisfies the discrete equations. These maps provide a qualitative comparison with the continuous analytical dispersion curves and the FFT spectra.
To enable direct comparison between the different diagnostics, the reference mesh used for the plane-wave residual analysis is scaled to a normalized grid spacing of $\Delta=k_0h=0.54$, matching the effective resolution of the $M_{\mathrm{15k}}$ mesh. For triangular elements, the plane-wave residual analysis uses a regular lattice of equilateral triangles, with the orientation shown in \cref{fig:3_mesh_Dofs} (\cref{subsec3_NUM_DISP}), whereas the FE simulations use irregular triangular meshes. The resulting maps therefore provide a structured reference for interpreting the FFT spectra, rather than an average dispersion response of the simulation meshes.

Whereas the plane-wave residual analysis probes the operator response to interpolated plane waves, spectral analysis through FFT evaluates the computed electric fields on the plasma side.
For the FFT diagnostic, simulations are performed with an enlarged plasma region, providing a longer spatial observation window and improving wavenumber resolution, while retaining the characteristic element size of the reference $M_\mathrm{15k}$ mesh.

\begin{figure}[pos=h!]
    \centering
    \includegraphics[width=1\linewidth]{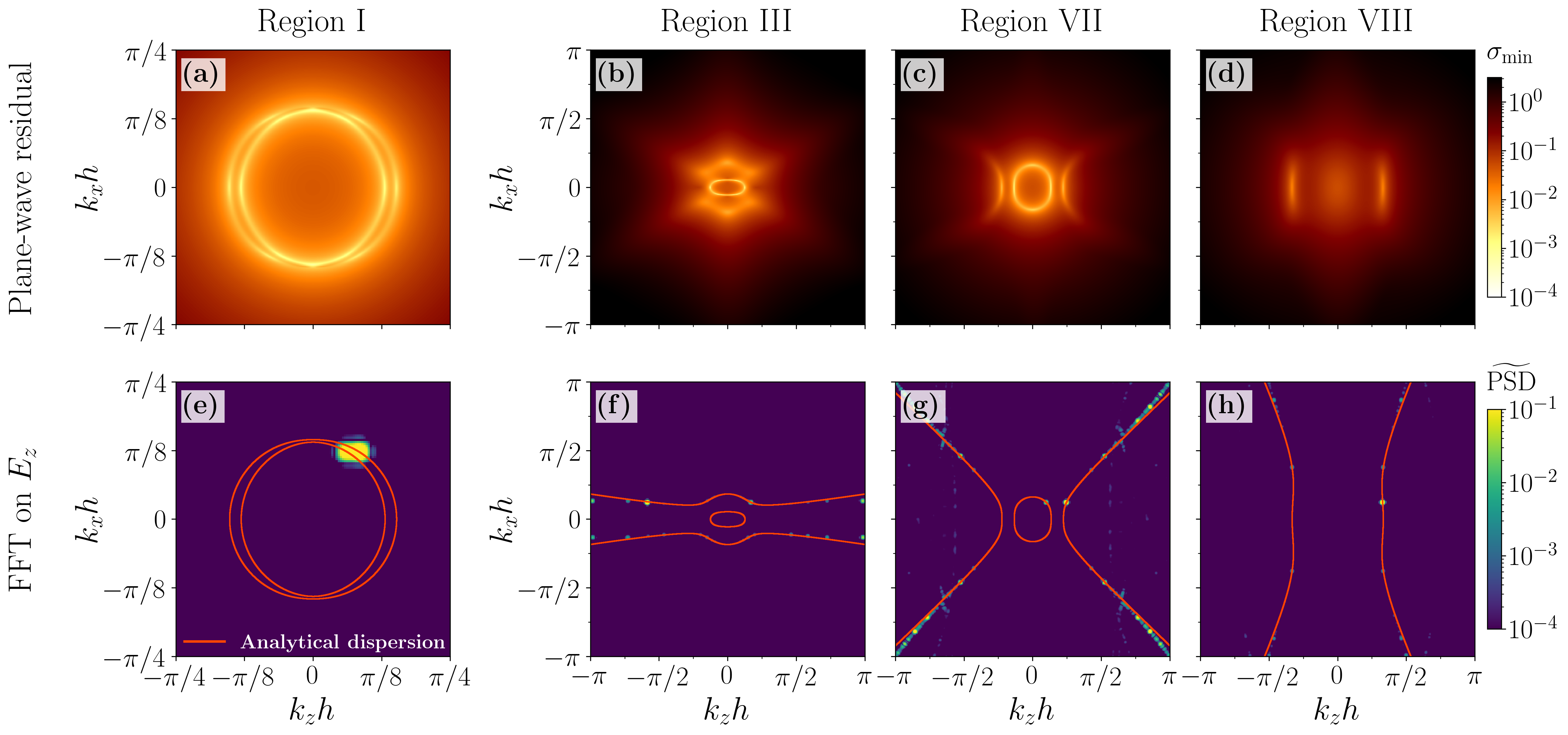}
    \caption{Spectral analysis of the second-order triangular discretization $\mathcal{T}^2$, comparing plane-wave residual maps with the Fourier spectra of the computed electric field. The top row [panels (a)--(d)] displays the smallest singular value $\sigma_{\min}$ as a function of the dimensionless wavenumbers $(k_zh,k_xh)$. Both rows cover $[-\pi,\pi]^2$, except for Region I [panels (a),(e)], where $[-\pi/4,\pi/4]^2$ is shown for clarity. The reference-mesh spacing is fixed at $\Delta=k_0h=0.54$, matching the effective spacing of the $M_{\text{15k}}$ mesh. The bottom row [panels (e)--(h)] shows the normalized PSD obtained from the two-dimensional FFT of the numerical axial field $E_z$. Each spectrum is normalized by its own maximum. Red curves overlaying the FFT spectra denote the exact continuous dispersion relations. Columns correspond to CMA Regions I, III, VII, and VIII, with the latter three supporting resonance cones.}
    \label{FIG4_dispersion_FFT_TRIS}
\end{figure}

The simulated axial field $E_z$ is first multiplied by a two-dimensional Hann window to reduce spectral leakage. The windowed field is zero-padded before the two-dimensional FFT to sample wavevector space more densely without changing the spectral content. The resulting Power Spectral Density (PSD) is normalized by its maximum value before being mapped into the same dimensionless wavevector space $(k_z h,k_x h)$, to identify the spurious spectral components introduced by the spatial discretization.

\cref{FIG4_dispersion_FFT_TRIS} presents complementary plane-wave residual and FFT diagnostics for the second-order triangular discretization $\mathcal{T}^2$ of the standard wave formulation. The top row [panels (a)--(d)] shows $\sigma_{\min}$ over the dimensionless wavevector domain. 
The bottom row [panels (e)--(h)] then displays the corresponding normalized PSD, overlaid with the exact continuous analytical dispersion curves as solid red lines.

In Region I, the low-residual contours in \cref{FIG4_dispersion_FFT_TRIS}(a) form a closed curve that follows the overall shape of the continuous dispersion relation shown in panel (e). The variation of $\sigma_{\min}$ along these contours indicates a directional dependence of the plane-wave residual on the reference triangular lattice.
This anisotropy does not lead to a broad spectral response in the Region I configuration: the PSD remains strongly localized around the analytically predicted transmitted wavevectors [panel (e)].
In the resonance-cone regimes (Regions III, VII, and VIII), the low-residual contours [panels (b)--(d)] exhibit a stronger directional dependence.
The corresponding FFT spectra [panels (f)--(h)] contain additional components at wavevectors absent from the exact transmitted field.
These components predominantly populate the  dispersion branches over an extended range of wavenumbers. At higher wavenumbers, slight offsets become visible with respect to the analytical solution, particularly in Regions III and VII, although the spectral distribution continues to follow the overall shape of these branches.

The combined plane-wave residual and FFT analyses reveal complementary effects of the triangular discretization.
The periodic triangular stencil used for the residual analysis retains discrete translational symmetry but produces a direction-dependent response to interpolated plane waves, consistent with the directional dispersion errors reported in earlier FE analyses of Maxwell's equations on triangular meshes \cite{monk1994}. A uniform Cartesian stencil also admits discrete plane-wave modes in the homogeneous interior, although their dispersion differs from the continuous relation. 
The unstructured triangular mesh used for the FEM simulations introduces spatial variations in element size, shape, and orientation. Its response to a plane wave therefore varies with the local geometry, and these variations can couple the transmitted field to additional wavevector components. 

The consequences of such coupling depend on the governing operator. In Region I, the principal symbol of the reduced scalar equation remains nonzero and grows in magnitude with wavenumber in every real direction. The resulting ellipticity provides the usual continuous estimates controlling short spatial scales \cite{tayl91}. 
The directional phase errors associated with the triangular discretization are not accompanied by substantial high-wavenumber spectral content: the PSD in panel (e) remains localized near the analytical transmitted wavevectors, consistently with the elliptic character of this regime.
In resonance-cone regimes, the reduced scalar equation instead loses ellipticity: its principal symbol vanishes along characteristic directions, and the usual elliptic control of short spatial scales no longer applies. 
In the collisionless cold-plasma model considered here, the continuous dispersion branches approach the resonance-cone asymptotes as the wavenumber increases without bound, admitting arbitrarily short wavelengths.
A mesh of characteristic size $h$ cannot resolve arbitrarily short wavelengths. Additional components with wavelengths of order $h$ can produce rapid variations between neighboring elements, providing a qualitative connection between the high-wavenumber FFT content and the spurious element-scale oscillations observed in \cref{FIG4_R8_2Dmaps}. 

\subsubsection{Uniform rectangular meshes}

\begin{figure}[pos=h!]
    \centering
    \includegraphics[width=1\linewidth]{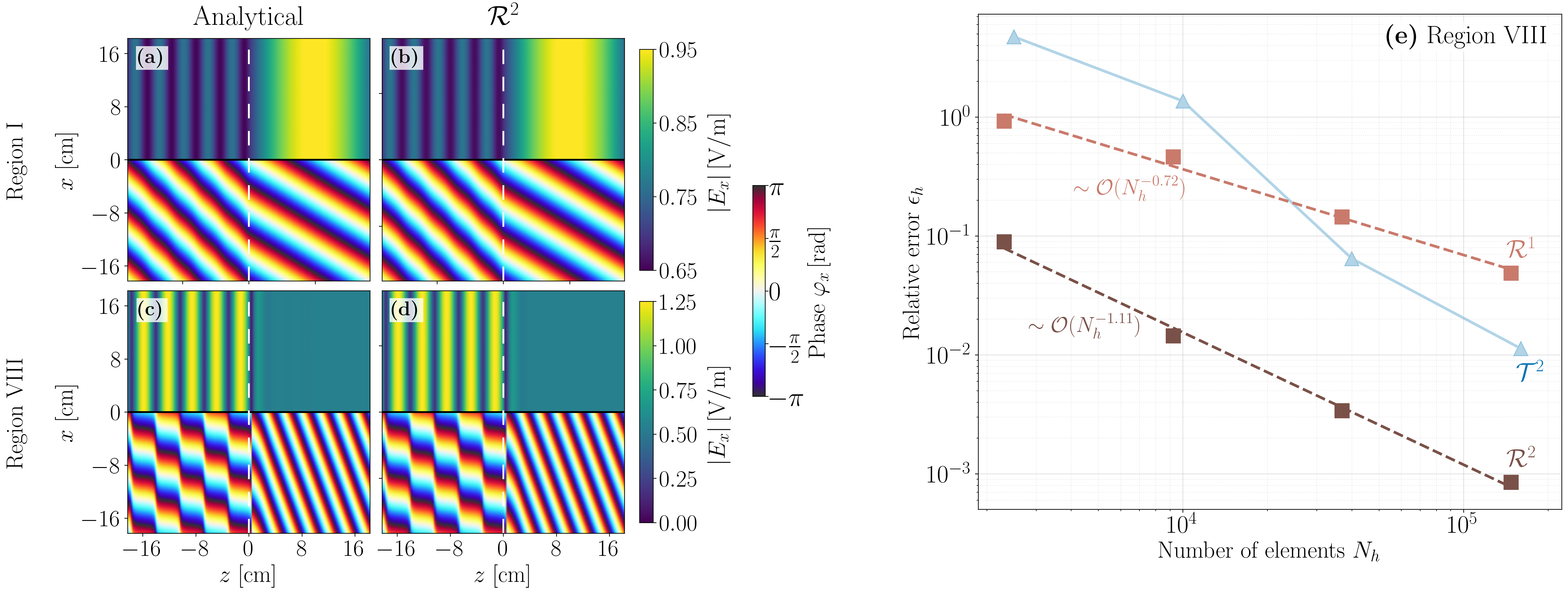}
    \caption{Spatial distributions of the $E_x$ component (left) and convergence analysis (right) for the rectangular discretizations $\mathcal{R}^1$ and $\mathcal{R}^2$. Panels (a),(c) show the analytical reference fields, while panels (b),(d) show the corresponding $\mathcal{R}^2$ numerical solutions for Region I (top) and Region VIII (bottom). The color scales are defined from the analytical solution in each region and applied unchanged to the numerical results. Panel (e) shows the relative error in Region VIII, as a function of the number of elements $N_h$ for $\mathcal{R}^1$ and $\mathcal{R}^2$. The collisionless second-order triangular results $\mathcal{T}^2$ are reproduced from \cref{FIG4_R8_conv_1} over the displayed range for comparison. Dashed lines indicate power-law fits for the rectangular discretizations.}
    \label{FIG4_RECT_convergence_maps}
\end{figure}

To assess the influence of mesh structure on the spurious oscillations observed with triangular elements, we now consider Cartesian grids composed of uniform rectangular elements.
\cref{FIG4_RECT_convergence_maps}(a)--(d) compares the analytical reference fields with the second-order rectangular solutions $\mathcal{R}^2$ of the standard wave formulation, computed on a mesh of comparable resolution to the $M_\mathrm{15k}$ triangular mesh. At the resolution of the figure, the numerical solution closely reproduces the analytical magnitude and phase distributions in both the elliptic Region I and the resonance cone regime of Region VIII. In particular, the element-scale oscillations observed with triangular discretizations are absent from the rectangular-grid solution [panel (d)].

The convergence analysis in \cref{FIG4_RECT_convergence_maps}(e) provides a quantitative assessment of this improvement in Region VIII. While the first-order rectangular discretization $\mathcal{R}^1$ mostly underperforms the second-order triangles $\mathcal{T}^2$ beyond sufficient mesh refinement, $\mathcal{R}^2$ achieves the highest accuracy over the complete range of tested resolutions.
Nevertheless, the comparison for a fixed number of elements $N_h$ requires some care. Rectangular elements contain more degrees of freedom than triangular elements of the same nominal polynomial order. Therefore, $N_h$ does not represent the same algebraic resolution for the two element families, and part of the horizontal separation between their convergence curves may arise from this difference. The present results demonstrate the accuracy obtained at fixed element count; however, a comparison at fixed number of degrees of freedom shows that $\mathcal{R}^2$ still clearly outperforms its triangular counterpart.
In addition, the relative error provides a global measure of accuracy but does not fully capture localized high-wavenumber pollution. The central result here is therefore that no spurious oscillations are observed with the rectangular discretizations in the collisionless regime, and the analytical wave structure is recovered.
Additional tests with $\mathcal{R}^2$ were performed by rotating the uniform background magnetic field within the propagation plane while keeping the Cartesian grid fixed. No spurious oscillations appeared over the range of orientations tested, showing that the favorable behavior observed in this configuration does not require alignment of the grid with $\bm{B}_0$.

\begin{figure}[pos=h!]
    \centering
    \includegraphics[width=0.95\linewidth]{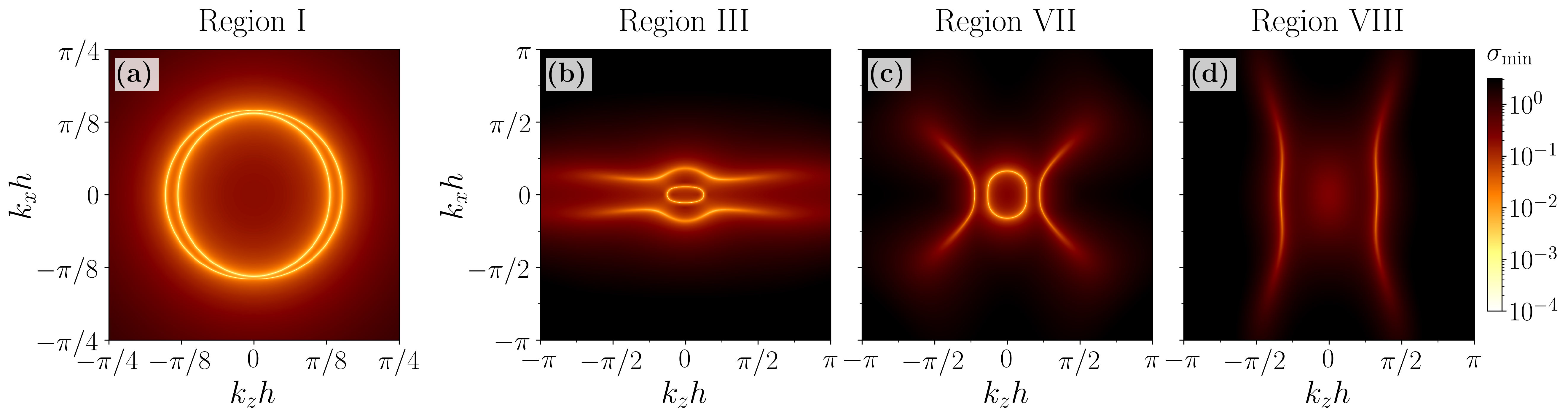}
    \caption{Plane-wave residual analysis of the second-order rectangular discretization $\mathcal{R}^2$, showing the smallest singular value $\sigma_{\min}$ of the reduced residual matrix as a function of the dimensionless wavenumbers $(k_zh,k_xh)$. The displayed domain spans $[-\pi,\pi]^2$, except in Region I [panel (a)], where $[-\pi/4,\pi/4]^2$ is shown for clarity. The normalized reference-grid spacing is fixed at $\Delta=k_0h=0.54$, matching the effective spacing of the $M_{15\mathrm{k}}$ reference mesh. Panels (a)--(d) correspond to CMA Regions I, III, VII, and VIII, with the latter three supporting resonance cones.}
    \label{FIG4_dispersion_RECT}
\end{figure}

The plane-wave residual maps in \cref{FIG4_dispersion_RECT} provide a complementary view of this behavior. Across the regimes examined, the low-residual contours reproduce the overall structure of the analytical dispersion branches, most clearly at small dimensionless wavenumbers.
In the resonance-cone regimes [panels (b)--(d)], the open branches follow the expected directions and are more clearly defined than in the corresponding triangular residual maps. 

The corresponding FFT spectra, not shown here, remain concentrated around the analytically predicted transmitted wavevectors and do not exhibit the extended population of the dispersion branches observed on the unstructured triangular meshes.
These observations are consistent with the spatial regularity of the Cartesian discretization. In the interior of each homogeneous region, the grid repeats the same local stencil. Numerical phase errors can still depend on propagation direction, but, for a given wavevector, the local discrete response is identical from cell to cell. 
This spatial regularity avoids the additional coupling between wavenumbers associated with variations in element geometry.

Small geometric perturbations of the Cartesian mesh reintroduce high-frequency oscillations in Region VIII. This was tested by introducing a sinusoidal deformation of the grid.
The absence of mesh-scale oscillations on the uniform grid is therefore fragile, not persisting on the distorted rectangular meshes tested here.
This requirement on mesh regularity is impossible to maintain in conforming discretizations of realistic device geometries, as their complex boundaries cannot be represented on a purely Cartesian grid.
Actual devices also include density and magnetic field gradients.
Furthermore, in 2D axisymmetric formulations, the cylindrical operator depends explicitly on $r$, breaking the translation invariance of the Cartesian problem and therefore altering its favorable dispersion properties.
These limitations motivate a regularization strategy compatible with unstructured meshes. Building on the improvement already observed in \cref{FIG4_R8_2Dmaps}, the next section examines the use of additional numerical collisionality to suppress spurious oscillations in resonance-cone regimes.

\subsection{On numerical collisionality} \label{subsec_4_res_coll}

\begin{figure}[pos=h!]
    \centering
    \includegraphics[width=1\linewidth]{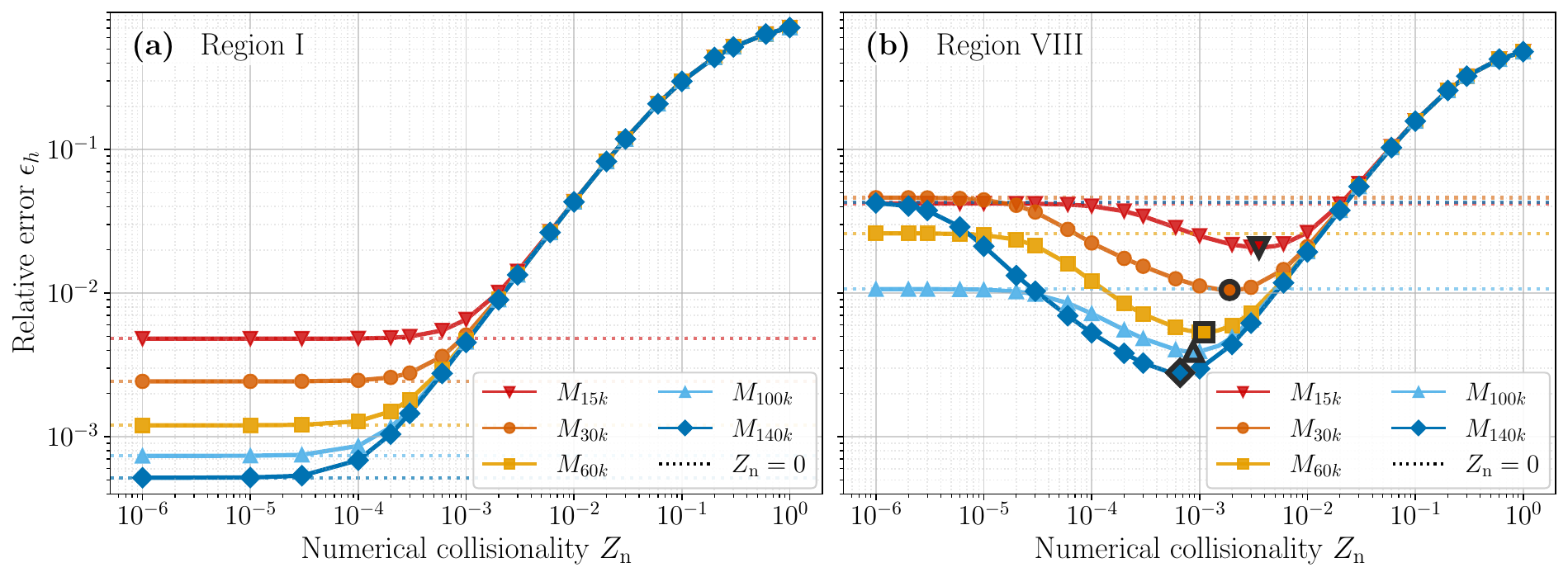}
    \caption{Effect of numerical collisionality $Z_\mathrm{n}$ on the relative error $\epsilon_h$ with respect to the collisionless analytical solution for $\mathcal{T}^2(Z_\mathrm{n})$, evaluated over progressively refined meshes. In Region I [panel (a)], the error remains close to its collisionless level at low collisionality before increasing at stronger damping. In Region VIII [panel (b)], an error-minimizing valley appears and shifts toward lower collisionality as the mesh is refined. For this figure, black marker outlines highlight the lowest sampled error on each curve. Horizontal dotted lines indicate the corresponding collisionless errors ($Z_\mathrm{n}=0$).}
    \label{FIG_err_vs_Z}
\end{figure}

The preceding results identify the second-order discretization $\mathcal{T}^2$ as the most suitable scheme among the triangular discretizations considered, combining reduced spurious oscillations with a favorable balance between accuracy and computational cost. Nevertheless, significant mesh-scale oscillations persist in collisionless resonance-cone regimes, indicating that neither mesh refinement nor higher-order approximation alone is sufficient to fully suppress the numerical pollution within computationally realistic resolutions. 
This motivates the introduction of numerical collisionality as a regularization mechanism, and this raises the question of how the required damping depends on mesh resolution. To address this point, \cref{FIG_err_vs_Z} extends the fixed-collisionality analysis of \cref{FIG4_R8_conv_1} by evaluating the relative error over a broad range of $Z_\mathrm{n}$ for several mesh sizes.

\cref{FIG_err_vs_Z}(a) shows the baseline behavior in the elliptic Region I, where numerical regularization is unnecessary. For sufficiently low values of $Z_\mathrm{n}$, the error remains stable, plateauing at a level mainly determined by the spatial resolution of the mesh. However, once $Z_\mathrm{n}$ exceeds a critical mesh-dependent threshold, the numerical damping error overtakes the spatial discretization error with respect to the benchmark collisionless solution, causing the total error to increase rapidly.
Beyond this threshold, the curves for different mesh resolutions merge onto a common rising branch, where the error is dominated by the difference between the continuous collisional and collisionless fields. Further refinement cannot remove this regularization error at fixed nonzero $Z_{\mathrm n}$.

\cref{FIG_err_vs_Z}(b) reveals different behavior in the resonance-cone Region VIII. At low collisionality, the error remains high, close to its collisionless level, reflecting the spurious oscillations observed on the unstructured triangular meshes. As $Z_{\mathrm n}$ is increased, the error decreases significantly, forming a distinct valley for each mesh resolution.
In this error-minimizing collisionality range, the damping is just strong enough to suppress the spurious oscillations, yet small enough to preserve the physical wave structure.
If $Z_\mathrm{n}$ is increased beyond this valley, the error increases, converging again to the damping-dominated trend with little dependence on mesh resolution, as for Region I. 
Additional collisionality limits wavenumber growth near resonance-cone directions and attenuates short-wavelength components, while also modifying the physical wave solution. The error minima in \cref{FIG_err_vs_Z}(b) show that the balance between suppressing numerical pollution and limiting these modifications is achieved at lower damping levels as the mesh is refined. A fixed value, such as $Z_{\mathrm n}=10^{-3}$, can therefore become excessive on finer meshes, where damping-induced changes increasingly limit accuracy. This explains the saturation observed with fixed additional collisionality in \cref{FIG4_R8_conv_1} and motivates a mesh-dependent choice.

\begin{figure}[pos=h!]
    \centering
    \includegraphics[width=0.5\linewidth]{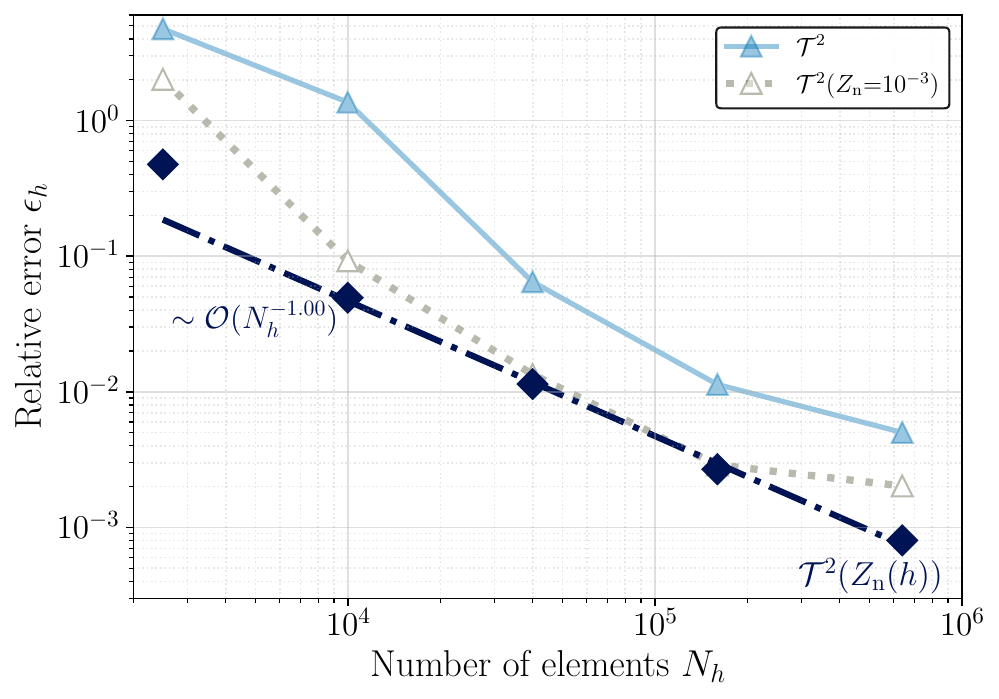}
    \caption{Convergence analysis in CMA Region VIII for the mesh-dependent regularization $\mathcal{T}^2(Z_{\mathrm n}(h))$. The relative error $\epsilon_h$, measured against the collisionless analytical solution, is plotted against the number of elements $N_h$. The collisionless $\mathcal{T}^2$ and fixed-collisionality $\mathcal{T}^2(Z_{\mathrm n}=10^{-3})$ curves are reproduced from \cref{FIG4_R8_conv_1} for comparison. Diamond markers show the mesh-dependent results. The dash-dotted line denotes a power-law fit, excluding the coarsest point, yielding $\epsilon_h\propto N_h^{-1.00}$.}
    \label{FIG4_R8_conv_2}
\end{figure}

To implement this adaptive scaling in this case, we extract the discrete error minima from the Region VIII analysis in \cref{FIG_err_vs_Z}(b). Fitting these optimal damping values against the corresponding mesh spacing $h$ gives an empirical power-law scaling $Z_\mathrm{n}(h)\approx10.2\,h^{1.48}$, with $h$ expressed in meters and a 95\% confidence interval of $1.21$--$1.76$ for the exponent. 
This relation is then used to repeat the convergence study of \cref{FIG4_R8_conv_1}, replacing the constant damping with this mesh-dependent collisionality. The results of this adaptive scheme, denoted as $\mathcal{T}^2(Z_\mathrm{n}(h))$, are presented in \cref{FIG4_R8_conv_2}.

The fitted damping law produces a sustained decrease in error as the mesh is refined. Excluding the coarsest point, the fitted dependence is $\epsilon_h\propto N_h^{-1.00}$, corresponding to $\epsilon_h\propto h^2$ for quasi-uniform two-dimensional meshes, recovering second-order convergence in this resonance-cone regime, while avoiding the saturation observed with fixed collisionality. 
This collisionality scaling also improves accuracy at mesh resolutions outside the range used for the fit, supporting its use for this benchmark.
The selected damping suppresses numerical pollution and improves agreement with the collisionless analytical reference, demonstrating an accuracy gain beyond field smoothing alone.

This empirical relation is calibrated for the homogeneous Region VIII configuration studied here and should not be transferred unchanged to other plasma conditions. In non-uniform plasmas, damping that improves the solution in resonance-cone regions can also modify propagation and absorption in non-resonant regions, as shown in \cref{FIG_err_vs_Z}(a), without providing an accuracy benefit. Its selection therefore requires a joint assessment of damping and spatial resolution, using analytical errors where available or relevant quantities of interest, as in the ECRT application considered next. Spatially varying collisionality offers a possible extension for concentrating regularization in regions where it is needed.
% ------------------------------------------------------

% ------------------------------------------------------
\section{Application to a waveguide-fed ECRT} \label{SEC5_ECRT}
The spurious oscillations investigated throughout the preceding section were encountered in previous waveguide-ECRT simulations \cite{inch24b}.
We now apply the numerical framework to this practical application, which features a non-uniform axisymmetric plasma, and assess the sensitivity of the computed fields and absorbed power to mesh resolution and additional collisionality.

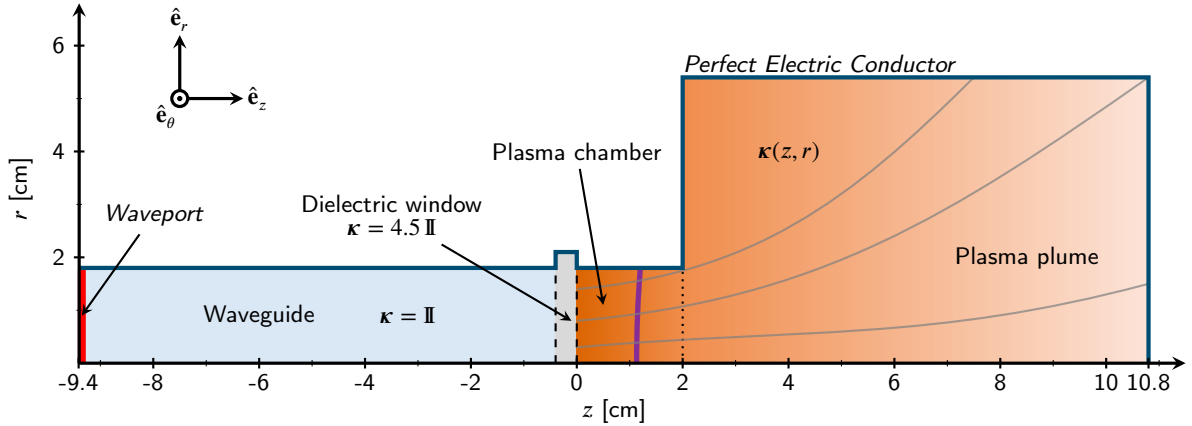
\begin{figure}[pos=h!]
    \centering
    \begin{tikzpicture}[x=1cm, y=1cm, scale=0.7]

    % Define custom colors
    \definecolor{waveguidebg}{RGB}{218, 232, 245}
    \definecolor{borderblue}{RGB}{0, 77, 115}
    \definecolor{ecrpurple}{RGB}{135, 45, 150}
    
    % Shading colors for continuous transition
    \definecolor{pcstart}{RGB}{220, 100, 0}
    \definecolor{pcend}{RGB}{240, 140, 75}
    \definecolor{plumend}{RGB}{252, 227, 215}

    % 1. Fill the background regions
    % Waveguide
    \fill[waveguidebg] (-9.4, 0) rectangle (-0.4, 1.8);
    % Dielectric window
    \fill[gray!30] (-0.4, 0) rectangle (0, 2.1);
    % Plasma Confinement (P.C.) - First part of shading
    \shade[left color=pcstart, right color=pcend] (0, 0) rectangle (2, 1.8);
    % Plasma plume - Second part of shading
    \shade[left color=pcend, right color=plumend] (2, 0) rectangle (10.8, 5.4);

    % 2. Draw interior boundary lines
    % Vertical dashed/dotted lines at specific coordinates
    \draw[thick, dashed] (-0.4, 0) -- (-0.4, 2.1);
    \draw[thick, dashed] (0, 0) -- (0, 2.1);
    \draw[thick, dotted] (2, 0) -- (2, 1.8);

    % ECR line (Purple curve)
    \draw[line width=2pt, ecrpurple] (1.13, 0) to[bend left=2] (1.2, 1.8);

    % Magnetic field line (Gray curve)
    \draw[thick, gray, opacity=0.7] (0, 0.3) to[out=5, in=195] (10.8, 1.5);
    
    \draw[thick, gray, opacity=0.7] (0, 0.8) to[out=5, in=215] (10.8, 5.4);
    
    \draw[thick, gray, opacity=0.7] (0, 1.4) to[out=5, in=225] (7.5, 5.4);

    % 3. Draw outer thick blue boundaries
    \draw[ultra thick, borderblue] 
        (-9.4, 1.8) -- 
        (-0.4, 1.8) -- 
        (-0.4, 2.1) -- 
        (0, 2.1) -- 
        (0, 1.8) -- 
        (2, 1.8) -- 
        (2, 5.4) -- 
        (10.8, 5.4) -- 
        (10.8, 0);

    % Waveport (Red line shifted slightly right to be clearly visible)
    \draw[line width=3pt, red] (-9.35, 0) -- (-9.35, 1.76);

    % 4. Draw coordinate axes and ticks
    % Z-axis
    \draw[very thick, ->, >=stealth] (-9.4, 0) -- (11.5, 0);
    % FIXED Z-AXIS LABEL: Manually shifted down to y = -0.8
    \node at (0.7, -0.9) {$z$ [cm]}; 
    
    % Z-axis Main Ticks (every 2cm)
    \foreach \x in {-8, -6, -4, -2, 0, 2, 4, 6, 8, 10} {
        \draw[thick] (\x, 0.1) -- (\x, -0.1) node[below, font=\small, inner sep=2pt] {\x};
    }
    
    % Z-axis Subticks (every 1cm, unlabelled, slightly shorter)
    \foreach \x in {-9, -7, -5, -3, -1, 1, 3, 5, 7, 9, 11} {
        \draw[thin] (\x, 0.06) -- (\x, -0.06);
    }
    
    % Z-axis Boundaries
    \draw[thick] (-9.41, 0.1) -- (-9.41, -0.1) node[below, font=\small, inner sep=2pt] {-9.4};
    \draw[thick] (10.8, 0.1) -- (10.8, -0.1) node[below, font=\small, inner sep=2pt] {10.8};

    % R-axis (placed at z = -9.4)
    \draw[very thick, ->, >=stealth] (-9.4, 0) -- (-9.4, 6.8);
    % FIXED R-AXIS LABEL: Manually shifted left to x = -10.2
    \node[rotate=90] at (-10.5, 3.3) {$r$ [cm]}; 
    
    % R-axis Main Ticks (every 2cm)
    \foreach \y in {2, 4, 6} {
        \draw[thick] (-9.3, \y) -- (-9.5, \y) node[left, font=\small] {\y};
    }
    
    % R-axis Subticks (every 1cm, unlabelled, slightly shorter)
    \foreach \y in {1, 3, 5} {
        \draw[thin] (-9.35, \y) -- (-9.45, \y);
    }

    % 5. Add Text Labels
    \node at (-6, 0.9) {Waveguide};
    \node at (-3.2, 0.9) {$\boldsymbol{\kappa} = {\rm I\! I}$};
    \node at (8.5, 2) {Plasma plume};
    \node at (4, 4) {$\boldsymbol{\kappa}(z, r)$};

    % Perfect Electric Conductor (Aligned left at z=2)
    \node[above right, font=\itshape, inner sep=0pt] at (2, 5.5) {Perfect Electric Conductor};

    % 6. Pointers and Annotations
    % Waveport pointer
    \node[font=\itshape] (waveport) at (-8, 2.8) {Waveport};
    \draw[->, >=stealth, thick] (waveport) -- (-9.35, 0.9);

    % Dielectric window pointer
    \node[align=center] (dieltext) at (-3.5, 2.8)
    {Dielectric window\\[-2pt]
    $\boldsymbol{\kappa} = 4.5\,{\rm I\! I}$};
    \draw[->, >=stealth, thick] (dieltext) -- (-0.1, 0.8);

    % % ECR line pointer
    % \node (ecrtext) at (0.5, 2.8) {ECR line};
    % \draw[->, >=stealth, thick] (ecrtext) -- (1.2, 1.1);
    
    % Discharge chamber pointer
    \node (chamber) at (0, 4) {Plasma chamber};
    \draw[->, >=stealth, thick] (chamber) -- (0.5, 1.1);

    % 7. Coordinate Triad (r, z, theta)
    \begin{scope}[shift={(-7.5, 5)}]
        % Theta component (circle with a dot)
        \draw[very thick] (0, 0) circle (0.15);
        \filldraw (0, 0) circle (0.04);
        \node[below left, xshift=2pt, yshift=1pt] {$\hat{\mathbf{e}}_\theta$};
        
        % Z and R unit vectors
        \draw[->, very thick, >=stealth] (0.15, 0) -- (1.2, 0) node[right, xshift=-2pt] {$\hat{\mathbf{e}}_z$};
        \draw[->, very thick, >=stealth] (0, 0.15) -- (0, 1.2) node[above, yshift=-2pt] {$\hat{\mathbf{e}}_r$};
    \end{scope}

\end{tikzpicture}
    \caption{2D axisymmetric simulation domain for full-wave ECRT modeling. The geometry is defined in cylindrical coordinates $(r,\theta,z)$ with the symmetry axis at $r=0$. 
    The thruster consists of a vacuum waveguide (light blue) of radius $1.8\,\mathrm{cm}$, a dielectric window of width $0.4\,\mathrm{cm}$ and relative permittivity $\varepsilon_r = 4.5$, and a plasma discharge chamber (dark orange) with its expanding plasma plume. 
    The nonuniform plasma density, collisionality, and applied magnetic field (represented by the grey streamlines) are read from a plasma transport simulation carried out with the HYPHEN code \cite{inch24b}. Microwave power is injected at the left waveport (red boundary). The top and right boundaries are PEC (bold blue lines).}
    \label{fig:5_simu_domain_ecrt}
\end{figure}

The waveguide-ECRT configuration of \cite{inch23a, inchingolo24, inch24b}, depicted in \cref{fig:5_simu_domain_ecrt}, is considered. The computational domain includes a vacuum waveguide, a dielectric window, a dense non-uniform magnetized plasma discharge and a lower-density plasma plume. 
The plasma properties are obtained from a transport simulation performed with the HYPHEN code for a radial-injection ECRT \cite{inch24b}, and define the spatially varying, anisotropic permittivity tensor $\boldsymbol{\kappa}(z,r)$.
Based on the comparisons in \cref{SEC4_Fresnel}, the ECRT simulations use the second-order triangular discretization $\mathcal{T}^2$.

The microwave power is injected at the left boundary of the vacuum waveguide through a waveport condition. We again use the formulation developed in \cref{subsec3_BC}, but in contrast to the 2D planar waveport, the ECRT application uses a circular waveguide. 
The simulations are performed at a frequency of $5.8\,\mathrm{GHz}$, for which only the fundamental $\mathrm{TE}_{11}$ mode propagates in the waveguide. Its transverse electric-field profile is defined analytically in terms of Bessel functions \cite{pozar11}. 
The linearly polarized mode can be reconstructed by superimposing the azimuthal Fourier modes $m=+1$ and $m=-1$ at the injection. The two modes are simulated independently because the problem is linear and the background configuration is axisymmetric. In the following, only the results for mode $m=+1$ are shown.

\begin{figure}[pos=h!]
    \centering
    \includegraphics[width=0.5\linewidth]{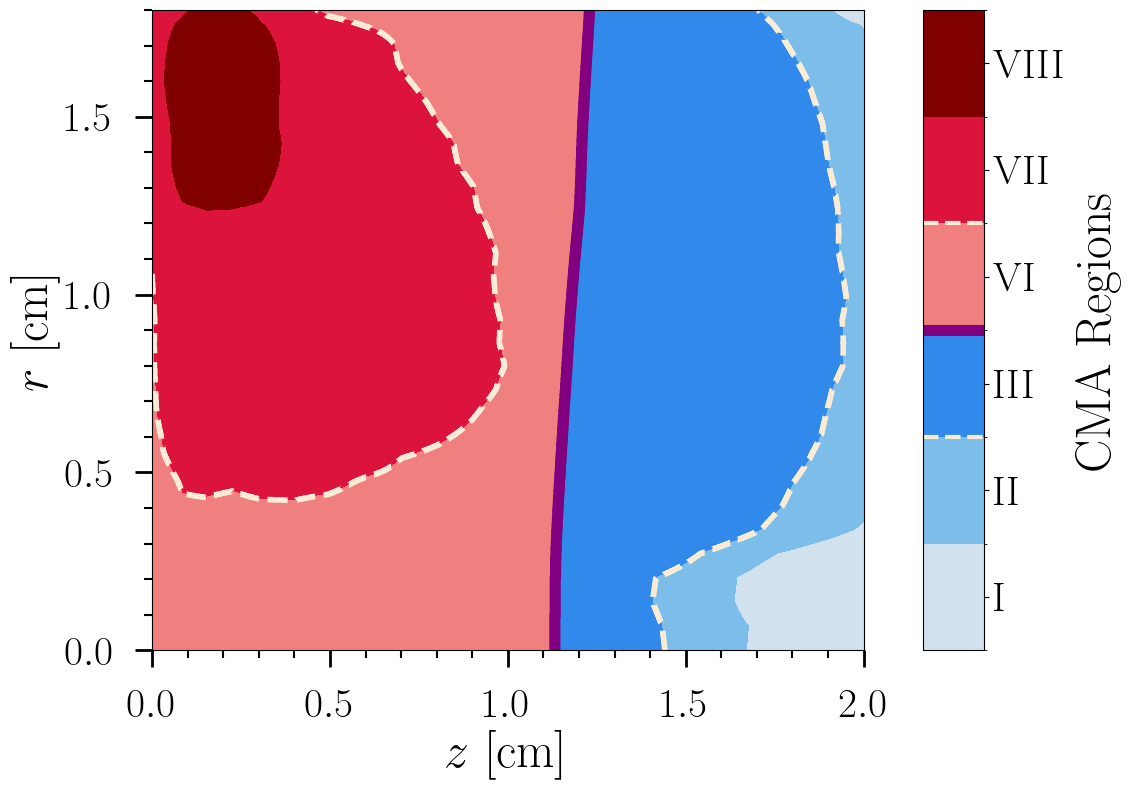}
    \caption{Topological map of the CMA regions within the waveguide-ECRT plasma discharge chamber (inner domain defined in \cref{fig:5_simu_domain_ecrt}). The plasma profile is derived from the steady-state maps for radial injection presented in \cite{inch24b}. The solid purple line identifies the ECR surface, and the white dashed lines mark boundaries between CMA regions with and without resonance cones.}
    \label{FIG5_regions_ECR}
\end{figure}

Based on these steady-state parameters, \cref{FIG5_regions_ECR} maps the CMA topological regions within the dense plasma discharge chamber. The map identifies the ECR surface and the fundamental transitions between different plasma regimes: the ordinary-mode cutoff (transition between Regions VII and VI), and the upper-hybrid resonance (transition between Regions III and II).
These boundaries are superimposed on the subsequent two-dimensional field and power maps in this section, to relate wave behavior and power absorption to the local plasma regime. As \cref{FIG5_regions_ECR} demonstrates, a substantial portion of the discharge chamber operates within resonance-cone regimes (Regions III, VII, and VIII). 

PEC boundary conditions are imposed at the downstream boundaries, following previous simulations \cite{jime23a,inch24b}, which are representative of the operation of this device within the walls of a laboratory vacuum chamber.
Nevertheless, the wave fields reaching the plume after passing through the resonance region are weak, so little power is reflected from the downstream walls. The simulation is therefore also representative of in-space operating conditions, where these walls are absent.
Given the weak fields in the plume, our numerical analysis focuses only on the discharge chamber.
The waveguide and plume regions are omitted from \cref{FIG5_regions_ECR} for visual clarity.

The different computational meshes used for the ECRT simulations are presented in \cref{tab:ecrt_mesh_params}, where the respective labels refer to the approximate number of elements in the plasma discharge chamber. 
The reported field and power-density maps in the rest of the section are restricted to this chamber, and integrated absorbed powers are evaluated over the same domain.

\begin{table}[pos=h!]
\centering
\begin{tabular}{ll@{\hspace{3em}}cccc@{\hspace{2em}}c}
\toprule
Parameter & Domain
& $M_{15\mathrm{k}}'$
& $M_{30\mathrm{k}}'$
& $M_{60\mathrm{k}}'$
& $M_{100\mathrm{k}}'$
& $M_{140\mathrm{k}}'$ \\
\midrule

\multirow{2}{*}{Number of elements $N_h$}
& Whole thruster
& 20,296
& 38,749
& 75,361
& 121,918
& 174,188 \\

& Discharge chamber
& 15,443
& 30,204
& 60,680
& 99,352
& 140,797 \\

\bottomrule
\end{tabular}
\caption{Spatial discretization parameters of the triangular meshes used for the waveguide-ECRT simulations: number of elements in the full computational domain and in the plasma discharge chamber.}
\label{tab:ecrt_mesh_params}
\end{table}

Unlike the collisionless Fresnel problem in \cref{SEC4_Fresnel}, the ECRT application includes some physical collisionality  $\nu_{\mathrm e}(\bm{r})$, supplied by HYPHEN. Adding numerical collisionality gives the effective collision frequency entering the dielectric tensor:
\begin{align}
    \nu(\bm{r}) = \nu_{\mathrm e}(\bm{r}) + Z_{\mathrm n}\omega.
\end{align}
Here, $Z_{\mathrm n}\omega$ represents the spatially uniform numerical contribution, and the total normalized collisionality is thus $Z(\bm{r})=\nu_{\mathrm e}(\bm{r})/\omega+Z_{\mathrm n}$. Setting $Z_{\mathrm n}=0$ therefore retains physical collisions while removing the additional numerical damping.

\subsection{Electric field structure and numerical regularization}

\begin{figure}[pos=h!]
    \centering
    \includegraphics[width=1\linewidth]{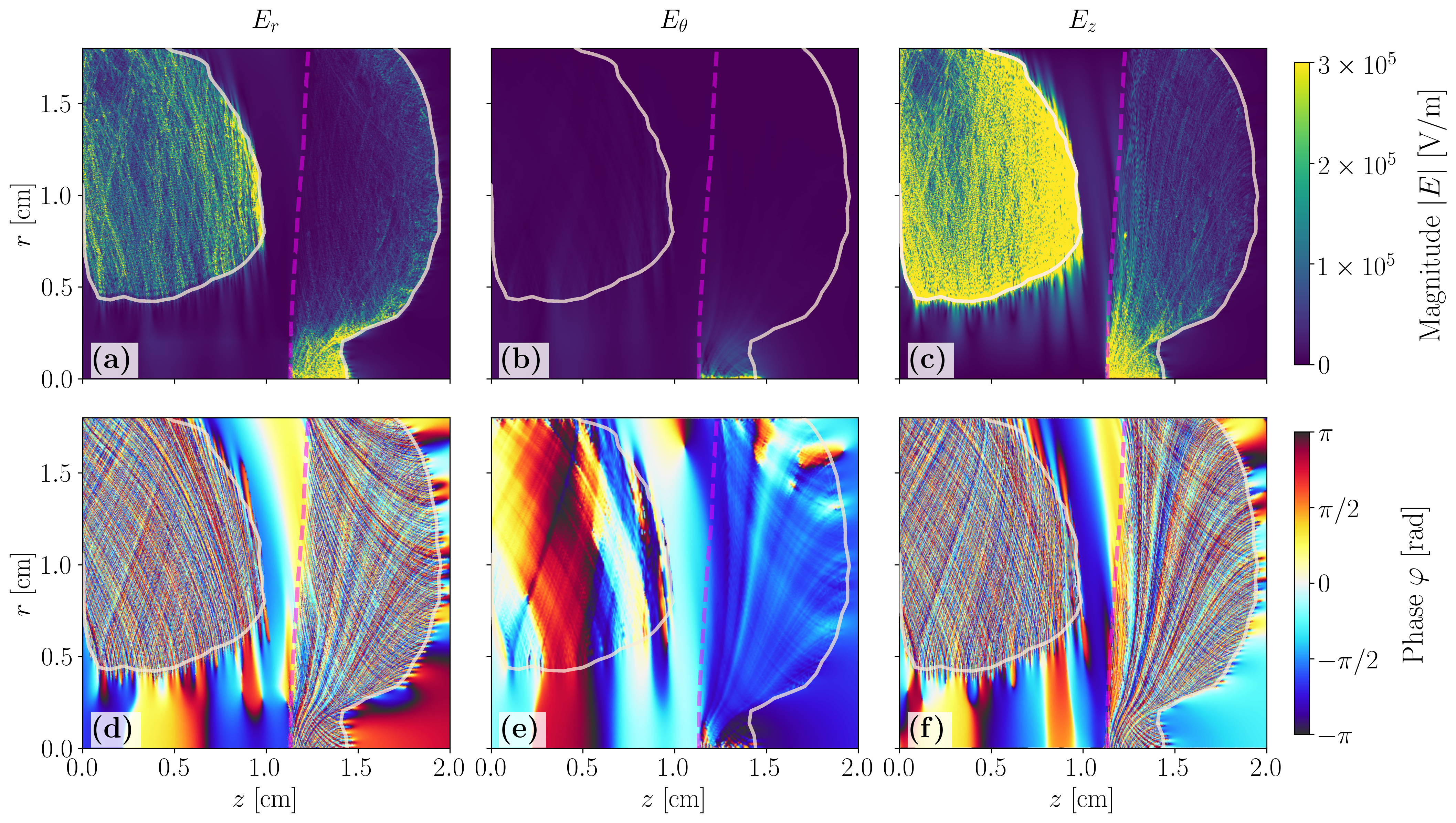}
    \caption{Spatial distributions of the magnitude [panels (a)--(c)] and phase [panels (d)--(f)] of the cylindrical electric field components ($E_r, E_\theta, E_z$, respectively) within the waveguide-ECRT discharge chamber, for $m = +1$. The numerical solution is obtained using the $\mathcal{T}^2$ discretization on the $M_{140\mathrm{k}}'$ mesh. This baseline case features no additional artificial collisionality ($Z_\mathrm{n}=0$). The dashed purple line identifies the ECR surface, and the white contours mark boundaries between CMA regions with and without resonance cones. Spurious oscillations in the in-plane fields $E_r$ and $E_z$ occur in regions that locally support resonance cones and are particularly evident in the phase maps.}
    \label{FIG5_fields_maps_Z=0}
\end{figure}

\cref{FIG5_fields_maps_Z=0} presents the magnitude and phase of the cylindrical electric field components ($E_r$, $E_\theta$, $E_z$) computed on the unstructured triangular mesh $M_{140\mathrm{k}}'$. 
For this baseline simulation, no additional numerical collisionality is applied ($Z_{\mathrm n} = 0$): the solver relies strictly on the physical electron collision frequency $\nu_\textrm{e}$.
The magnitudes correspond to a total absorbed power of $100\,\mathrm{W}$ in the discharge chamber.

Despite physical collisional damping, the numerical solution exhibits short-wavelength spurious oscillations similar to those found in \cref{SEC4_Fresnel}. The phase maps [panels (d),(f)] show strong oscillations dominating the in-plane components (Nédélec space), while the out-of-plane component has a smoother behavior.
These artifacts are confined to regions that locally support resonance cones in the collisionless cold-plasma model, as indicated by the superimposed CMA boundaries.

This spatial correspondence is consistent with the generalized Fresnel benchmark, where the reduced scalar equation loses ellipticity in these regimes. 
At the mesh resolution considered, the physical collision frequency $\nu_\mathrm{e}(\bm{r})$ alone is  insufficient to suppress this numerical pollution. 
For reference, its normalized value peaks near the gas radial injection site, at the top left corner of the domain, at only $\nu_\mathrm{e,max}/\omega = 4.93 \times 10^{-4}$, and decreases by two orders of magnitude toward the limits of the discharge chamber boundaries.
Across the tested range of mesh resolutions, refinement does not substantially reduce these oscillations. Instead, their spatial scale decreases with the element size, consistent with their mesh-scale character.

\begin{figure}[pos=h!]
    \centering
    \includegraphics[width=1\linewidth]{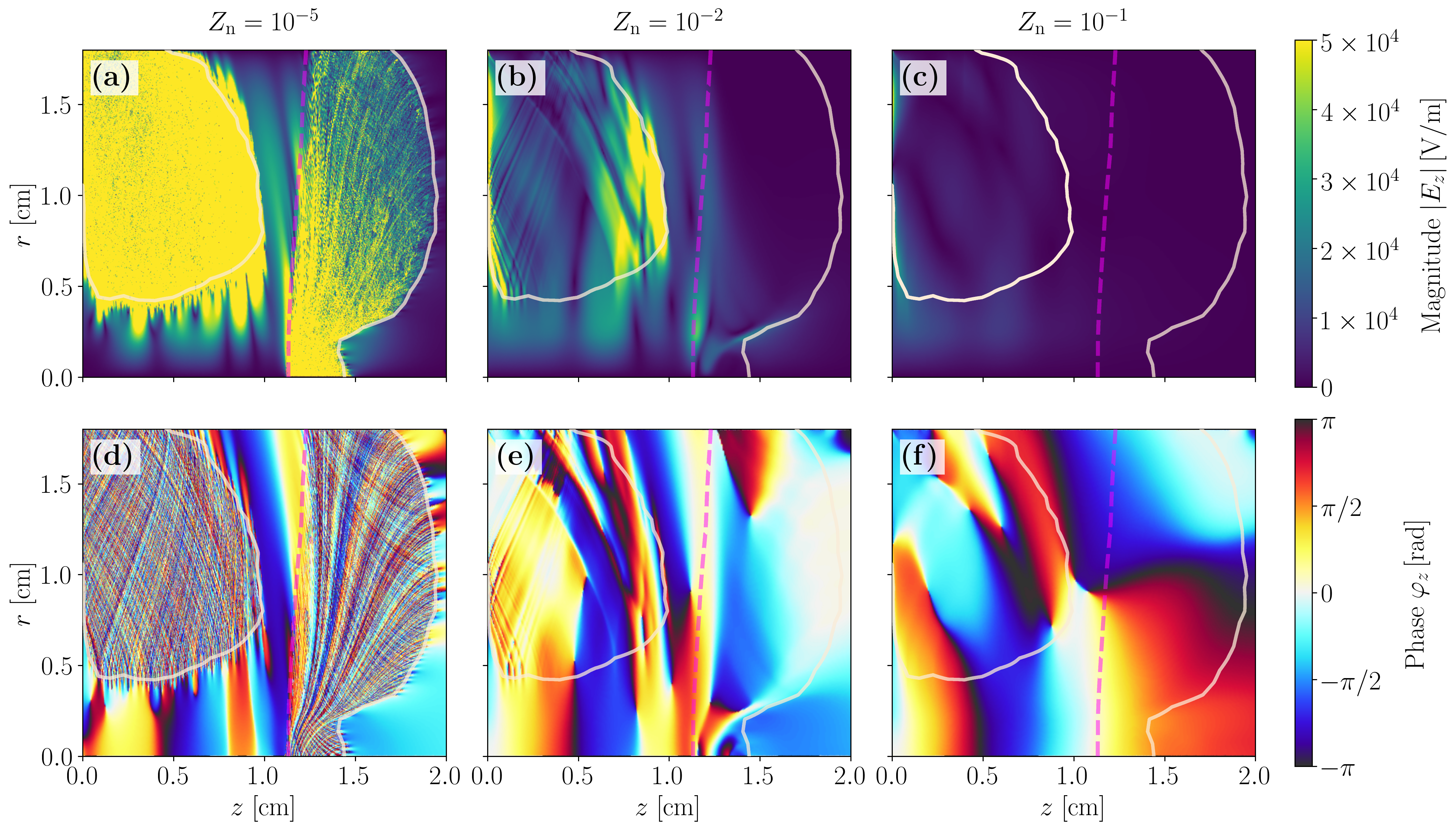}
    \caption{Spatial distributions of the magnitude [panels (a)--(c)] and phase [panels (d)--(f)] of the axial electric field component ($E_z$) within the waveguide-ECRT discharge chamber, for $m=+1$. The columns correspond to increasing levels of artificial collisionality: $Z_{\mathrm n} = 10^{-5}$ (left), $Z_{\mathrm n} = 10^{-2}$ (center), and $Z_{\mathrm n} = 10^{-1}$ (right). As with the baseline case in \cref{FIG5_fields_maps_Z=0}, the numerical solution is obtained using the $\mathcal{T}^2$ discretization on the  $M_{140\mathrm{k}}'$ mesh. Consistent with the topological map established in \cref{FIG5_regions_ECR}, the dashed purple line identifies the ECR surface, and the white contours mark boundaries between CMA regions with and without resonance cones.}
    \label{FIG5_Ez_map_collisions}
\end{figure}

To reduce these resolution requirements, we introduce additional artificial collisionality through the parameter $Z_{\mathrm n}$.
In \cref{FIG5_Ez_map_collisions} we compare the axial electric field magnitude and phase for three representative values of the numerical collisionality $Z_{\mathrm n}$. The magnitudes correspond to the power absorption density shown later in \cref{FIG5_Qa_map_collisions}. 
For low collisionality, $Z_{\mathrm n} = 10^{-5}$ [panels (a),(d)], the solution remains noisy. Severe spurious oscillations continue to dominate the resonant regions, closely resembling the unregularized baseline in \cref{FIG5_fields_maps_Z=0}(c),(f).

Conversely, for high collisionality, $Z_{\mathrm n} = 10^{-1}$ [panels (c),(f)], the solution becomes overdamped. The mesh-scale oscillations observed at low collisionality are completely gone from the phase map, which instead exhibits smooth variations on larger spatial scales.
However, the field magnitude is strongly reduced and barely visible on the common color scale, while the phase pattern differs markedly from the intermediate-collisionality case. Thus, the high damping level suppresses the observed numerical pollution while substantially modifying the computed wave response.

At the intermediate value $Z_{\mathrm n}=10^{-2}$, most mesh-scale oscillations are suppressed and more coherent phase fronts are obtained. 
Some oblique structures remain, with characteristic spatial scales larger than the local element size and stable over mesh refinement. 
This suggests those remaining structures are likely physical in origin.

\subsection{Mesh and damping sensitivity of power absorption}

In most applications, the microwave power coupled into the plasma is the key output of wave simulations for EPTs. 
In \cref{SEC4_Fresnel}, the influence of the additional collisionality introduced for numerical regularization was assessed through the error relative to the analytical field solution. Here, however, its ability to reduce spurious oscillations must also be assessed in terms of the deposited power, since power absorption depends directly on the total damping. In this section we therefore examine the sensitivity of the deposited power to both mesh resolution and additional collisionality $Z_\mathrm{n}$.
The local time-averaged absorbed power density $Q_\textrm{a}$ is determined by:
\begin{equation}
    Q_\textrm{a}(z, r) = \frac{1}{2} \text{Re} \left( \bm{J}^* \cdot \bm{E} \right)
\end{equation}
with $\bm{J} = i \omega \varepsilon_0 (\id - \bm{\kappa}) \bm{E}$ the induced plasma current density. 
For consistency, the power absorption $Q_\textrm{a}$ is computed in post-processing using the same regularized tensor $\bm{\kappa}$ (including artificial collisionality) that is used in the weak formulation to solve for the electric fields.  
The power absorbed in the discharge chamber is integrated over the volume of the plasma discharge chamber $\Omega$. Taking into account the 2D axisymmetric geometry, the total absorbed power $P_\textrm{a}$ is evaluated as:
\begin{equation}
    P_\textrm{a} = \int_{\Omega} Q_\textrm{a}(z, r) \, 2\pi r \, \:\mathrm{d}z \:\mathrm{d}r.
\end{equation}
Exploiting the linearity of the wave equation, simulations are first performed with an arbitrary excitation amplitude. Then, for a standardized comparison across collisionality regimes, all power density maps $Q_\textrm{a}(z,r)$ are multiplied by the same factor $P_\textrm{target}/P_{\textrm{a,ref}}$, where $P_\textrm{target}=100\,\mathrm{W}$ and $P_{\textrm{a,ref}}$ is the integrated absorbed power of the reference case ($Z_{\mathrm n}=10^{-2}$ on $M_{140\mathrm{k}}'$). This common normalization thus sets the reference absorbed power to $100\,\mathrm{W}$ while preserving the relative differences between cases.

\begin{figure}[pos=h!]
    \centering
    \includegraphics[width=1\linewidth]{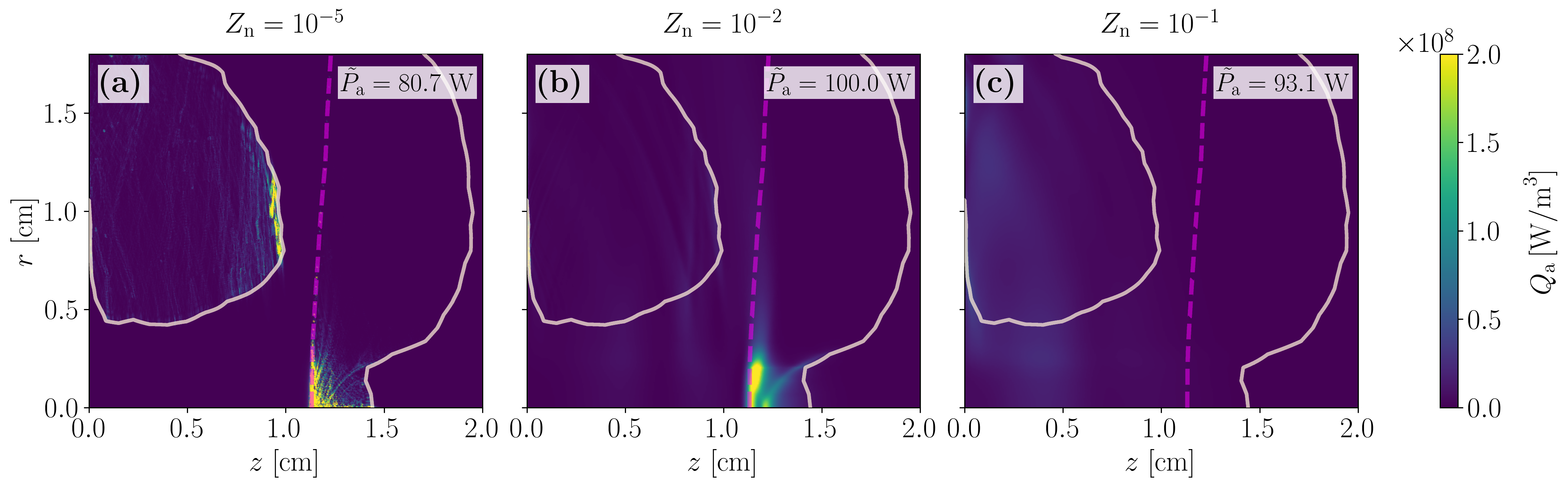}
    \caption{Spatial distributions of the power absorption density $Q_\textrm{a}$ within the waveguide-ECRT discharge chamber, for $m=+1$. The panels correspond to increasing levels of artificial collisionality: $Z_{\mathrm n} = 10^{-5}$ (a), $Z_{\mathrm n} = 10^{-2}$ (b), and $Z_{\mathrm n} = 10^{-1}$ (c), respectively. The inset values in the top right corners denote the rescaled integrated absorbed power $\tilde{P_\textrm{a}}$. Consistent with the topological map established in \cref{FIG5_regions_ECR}, the dashed purple line identifies the ECR surface, and the white contours mark boundaries between CMA regions with and without resonance cones.}
    \label{FIG5_Qa_map_collisions}
\end{figure}

\cref{FIG5_Qa_map_collisions} presents the resulting spatial distributions of the time-averaged power absorption density across the three collisionality regimes previously analyzed in \cref{FIG5_Ez_map_collisions}. The values $\tilde{P}_\textrm{a}$ indicated in each panel give the corresponding integrated absorbed powers after applying this global normalization.
Both $P_\textrm{a}$ and $\tilde{P}_\textrm{a}$ refer exclusively to absorption by mode $m=+1$ within the discharge chamber.
These maps demonstrate that the numerical collisionality $Z_{\mathrm n}$ affects both the spatial localization of power deposition and the overall wave coupling efficiency.

In the low-collisionality regime, $Z_{\mathrm n}=10^{-5}$ [panel (a)], power absorption is concentrated around the ECR surface, as expected, but additional absorption also appears along the boundary between Regions VII and VI. The artificial collisionality is insufficient to regularize the fields.
As a result, sparse spurious peaks can still be identified along the ECR layer upon closer inspection. This case yields the lowest integrated power ($\tilde{P}_a = 80.7\,\mathrm{W}$).
The overdamped regime, $Z_{\mathrm n}=10^{-1}$ [panel (c)], exhibits a diffuse absorption profile extending upstream of the ECR surface, with strongly reduced absorption at the resonance. 
This excessive damping leads to an increased reflected power,  yielding a rescaled integrated absorbed power of $\tilde{P}_\textrm{a}=93.1\,\mathrm{W}$.
In the intermediate case, $Z_{\mathrm n} = 10^{-2}$ [panel (b)], absorption is concentrated near the ECR surface, particularly near the axis, with a smoother spatial distribution.
A fraction of the power is also coupled toward the upper-hybrid resonance, indicated by the white boundary separating Regions III and II on the right side of the domain.

\begin{figure}[pos=h!]
    \centering
    \includegraphics[width=0.65\linewidth]{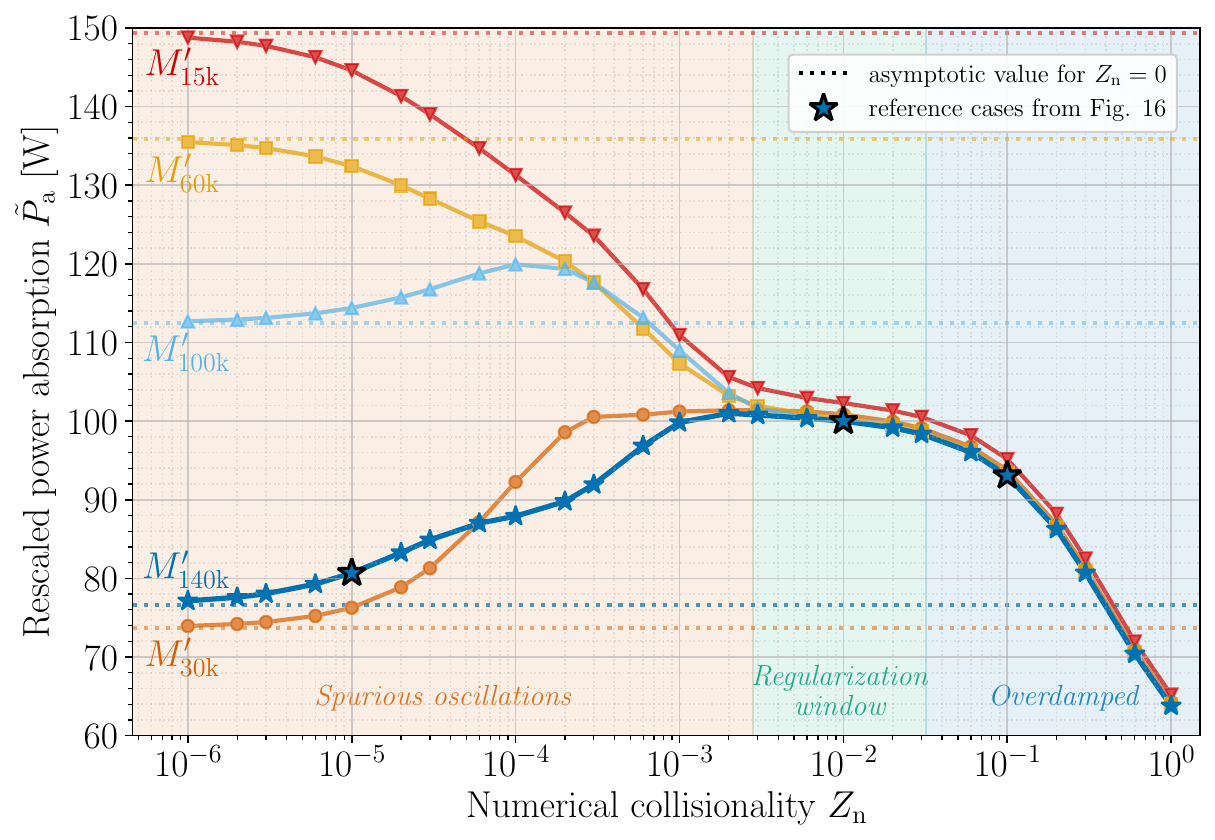}
    \caption{Evolution of the rescaled integrated absorbed power $\tilde{P}_a$ as a function of the numerical collisionality $Z_{\mathrm n}$ within the waveguide-ECRT discharge chamber, for $m=+1$. The numerical solution is obtained using the $\mathcal{T}^2$ discretization on meshes specifically refined within the discharge chamber. The power values are rescaled such that the reference configuration, the $M_{140\mathrm{k}}'$ mesh at $Z_{\mathrm n} = 10^{-2}$, equals exactly $100\,\mathrm{W}$. The parametric sweep compares five distinct mesh resolutions ($M_{15\mathrm{k}}'$ to $M_{140\mathrm{k}}'$), showing three numerical regimes: a low-collisionality region (orange background) dominated by spurious field oscillations; a regularization window (green background) characterized by a plateau, with stable power values; and an overdamped region (blue background) where power absorption decreases sharply. Horizontal dotted lines denote the asymptotic power limits for $Z_{\mathrm n} = 0$, using only the baseline electron collision frequency provided by the transport model. Star markers highlight the specific reference cases shown as power absorption density maps in \cref{FIG5_Qa_map_collisions}.}
    \label{FIG5_Pa_vs_Z_ECRT}
\end{figure}

\cref{FIG5_Pa_vs_Z_ECRT} presents the evolution of the rescaled integrated absorbed power $\tilde{P}_\mathrm{a}$ against $Z_{\mathrm n}$ for five distinct mesh resolutions. The common normalization uses the reference case $Z_{\mathrm n}=10^{-2}$ on the $M_{140\mathrm{k}}'$ mesh, shown in \cref{FIG5_Qa_map_collisions}(b). 
The parametric sweep identifies three damping regimes.
In the low-collisionality regime ($Z_{\mathrm n}<3\times10^{-3}$), spurious oscillations persist and the integrated absorbed power depends strongly and non-monotonically on mesh resolution. Across the tested grids, it exceeds the reference value by up to $50\%$ or falls below it by $26\%$. This sensitivity limits the reliability of the predicted wave-plasma power coupling and further motivates introducing numerical collisionality to obtain more consistent absorbed-power estimates across mesh resolutions.
In the high-collisionality limit ($Z_{\mathrm n} > 3\times10^{-2}$), the absorbed power decreases monotonically, with the same trend across all the tested mesh resolutions.
In this overdamped regime, excessive artificial collisionality dominates wave propagation, and makes the plasma more reflective. This reduces wave-plasma coupling while redistributing the remaining absorption upstream, as illustrated in \cref{FIG5_Qa_map_collisions}(c).

Between these two extremes, for $Z_{\mathrm n} \in [3 \times 10^{-3}, 3 \times 10^{-2}]$, there exists a regularization window where the power curves for all meshes form a common plateau. The absorbed power varies only weakly with damping and mesh resolution: although finer meshes yield slightly lower power values, the relative difference across the wide range of resolutions studied remains below $4\%$. This weak dependence on damping is consistent with the result of Després et al. \cite{despres2014}, who showed that absorption at a cold plasma hybrid resonance can approach a finite, nonzero limit as collisionality tends to zero. 
The close agreement between meshes supports estimating the absorbed power within this plateau for the configuration considered here.

The collisionality values within this plateau lie below the reference value $\nu_\mathrm{eff}/\omega\simeq0.05$ discussed by Hagelaar et al. \cite{hage09} for their ECR model. Below this level, they reported substantially improved numerical convergence without significant changes in the overall power profile. The present results exhibit similar qualitative behavior.
More directly comparable to the present formulation, Yasaka et al. \cite{yasaka92} used normalized effective collision frequencies of $10^{-3}$--$10^{-2}$ in 2D FEM simulations of ECR plasmas, overlapping the plateau we observe.
A similar collisionality was also required in the second-order FEM simulations of Otin et al. \cite{otin26}, whose reported numerical stability threshold of $\nu=10^9\,\mathrm{s}^{-1}$ at a wave frequency of $28\,\mathrm{GHz}$ corresponds to $\nu/\omega\simeq5.7\times10^{-3}$.
Accordingly, $Z_{\mathrm n}=10^{-2}$ is adopted for the reference results in \cref{FIG5_Ez_map_collisions,FIG5_Qa_map_collisions}. This value lies well within this plateau, suppressing most spurious oscillations while limiting artificial attenuation upstream of the ECR surface. The appropriate damping level, however, remains dependent on the specific plasma configuration.

\paragraph{Impact on computational cost --- }

An additional observation concerns the influence of numerical collisionality on computational time. On the finest mesh $M_{140\mathrm{k}}'$, the time required by $\mathcal{T}_\mathrm{W}^2$ decreases by more than an order of magnitude over the investigated range of $Z_{\mathrm n}$, whereas $\mathcal{T}_\mathrm{SP}^2$ exhibits a much more moderate variation.
While $\mathcal{T}_\mathrm{SP}^2$ is faster at weak damping, $\mathcal{T}_\mathrm{W}^2$ becomes faster as $Z_{\mathrm n}$ increases. 
The crossover between the two formulations occurs within the absorbed-power plateau, where the computational time of $\mathcal{T}_\mathrm{W}^2$ decreases sharply and the absorbed power remains nearly constant. The relative computational cost can therefore vary substantially without any significant change in the power absorption.

Additional tests for $\mathcal{T}_\mathrm{W}^2$ show that the estimated matrix condition number generally decreases as $Z_{\mathrm n}$ increases and rises strongly with mesh refinement: the reduction in computational time accompanies an improvement in conditioning.
With the present solver, the most computationally efficient formulation depends on both mesh resolution and damping level. Nevertheless, the lower sensitivity of $\mathcal{T}_\mathrm{SP}^2$ to collisionality results in more predictable computational costs, which can represent a practical advantage for broader parametric studies.

\paragraph{Alternative implementation of numerical collisionality --- }

The spatial distribution of the additional collisionality can also influence the balance between numerical regularization and artificial attenuation.
In this work, we added a globally constant damping term scaled by the wave’s angular frequency, so that the total collisionality is $\nu(\bm{r}) = \nu_\mathrm{e}(\bm{r}) + Z_{\mathrm n}\omega$. This prescription uses one adjustable parameter, is easy to implement, and regularizes the solution when the damping is appropriately calibrated. 
During the development of this study, two alternative definitions were evaluated. The first approach consists of directly scaling the spatial electron collision frequency provided by the transport code via a constant multiplier, such that $\nu(\bm{r}) = \beta \nu_\mathrm{e}(\bm{r})$. 
The second alternative scales the numerical collision frequency with the local electron cyclotron frequency, using the anomalous transport coefficient $\alpha_\mathrm{t}$ employed in HYPHEN: $\nu(\bm{r}) = \nu_\mathrm{e}(\bm{r}) + \alpha_\mathrm{t}\omega_\mathrm{ce}(\bm{r})$. Retaining the value of $\alpha_\mathrm{t}$ used in the transport model avoids introducing an independently adjustable damping parameter in the wave solver.

Numerical tests using these alternative definitions of additional collisionality yielded spatial power-density distributions and integrated absorbed powers similar to those obtained with the uniform additive term presented in this section. 
This agreement suggests that, for this configuration, the wave–plasma coupling is relatively insensitive to the choice among the tested prescriptions, as long as a suitably calibrated damping level is defined.
The appropriate parameters should nevertheless be assessed separately through mesh and collisionality sensitivity studies, balancing suppression of spurious oscillations against overdamping and increased reflected power.
Alternatively, artificial damping could be localized to regions requiring regularization.

% ------------------------------------------------------

% ------------------------------------------------------
\section{Conclusion}\label{sec:conclu}
This work has investigated the origin of mesh-scale oscillations in FEM simulations of electromagnetic waves in magnetized cold plasmas and assessed numerical strategies for reducing them. 
A mixed Nédélec--Lagrange solver and a waveport condition accommodating non-orthogonal modes in anisotropic media provided a common framework for the planar Fresnel benchmark and the axisymmetric ECRT application. 
For the generalized Fresnel benchmark, we derived an exact analytical solution that enabled direct verification of the computed fields. This comparison established that the oscillations observed in resonance-cone regimes are numerical artifacts.

This numerical pollution remains strongly dependent on the wave propagation regime, mesh structure and approximation order. In Region I, the FFT spectra remain localized around the analytically predicted transmitted wavevectors, whereas in resonance-cone regimes, additional spectral components populate the analytical dispersion branches.
The observed sensitivity to mesh structure suggests that local element variations couple the transmitted field to additional wavevectors in regimes where the reduced scalar equation loses ellipticity.
Consistent with this interpretation, uniform rectangular grids avoid the observed spurious oscillations, including cases with oblique background magnetic fields, whereas small mesh deformations reintroduced them.

Among the tested triangular discretizations, second-order elements provided the most favorable accuracy-cost balance. For a given approximation order, $\mathrm{W}$ and $\mathrm{SP}$ produced matching fields and errors, as expected from their equivalence, and both formulations retained oscillations in Region VIII on unstructured triangular meshes.
At second order, the main advantage of $\mathrm{SP}$ was computational in our implementation: with the specific direct solver used here, lower LU fill-in improved its scaling performance on fine meshes in the planar benchmark. Its computational time was also less sensitive to collisionality in the ECRT calculations, although the faster formulation of the two depends strongly on mesh resolution and damping level.

The analytical Fresnel reference also allowed us to quantify the effect of additional collisionality on solution accuracy. Artificial damping provided no appreciable accuracy benefit in the elliptic Region I, where the numerical fields showed no spurious oscillations.
In contrast, an intermediate damping level reduced the error in the resonance-cone Region VIII, yielding a distinct error minimum over the tested collisionality range.
The error-minimizing numerical collisionality decreases with mesh refinement, explaining why a fixed nonzero damping level eventually limited accuracy relative to the collisionless reference. 
For this benchmark, the empirical scaling $Z_{\mathrm n}\propto h^{1.48}$ avoided the fixed-damping error plateau in Region VIII, and sustained error reduction consistent with second-order convergence over the tested resolution range.

For the non-uniform ECRT plasma, where no analytical reference solution is available, the study focused on power absorption. Physical collisions alone did not remove the mesh-scale oscillations at the tested resolutions, whereas excessive numerical collisionality altered wave propagation and increased reflected power.
Between these limits, the tested meshes demonstrate the existence of a common absorbed-power plateau for $Z_{\mathrm n}\in[3\times10^{-3},3\times10^{-2}]$. Within this interval, integrated power varied only weakly with damping, and differences between meshes remained below $4\%$. The plateau, together with the field and power-density maps, supports choosing $Z_{\mathrm n}=10^{-2}$ as the reference value for this configuration.

The generalized Fresnel benchmark and the waveguide-ECRT application provide complementary criteria for selecting numerical collisionality: field error and absorbed-power sensitivity. 
These results offer practical guidance for reducing numerical pollution in cold-plasma FEM simulations, with suitable damping levels depending on the configuration and discretization considered.
% ------------------------------------------------------

\section*{Acknowledgments}

The authors thank P. Jiménez for the insightful discussions on the previous version of the wave code, and M. R. Inchingolo for the waveguide-ECRT plasma density maps used for the wave simulations presented in \cref{SEC5_ECRT}.

This work has received funding from the European Research Council (ERC) under the European Union’s Horizon 2020 research and innovation programme (Starting Grant ZARATHUSTRA, grant agreement No 950466).

This work has been supported by the R\&D project PID2023-150052OB-I00 (ADAPT) funded by \\ 
MICIU/AEI/10.13039/501100011033 and by ERDF, EU.

The Department of Aerospace Engineering at Universidad Carlos III de Madrid has been accredited as a Unit of Excellence “María de Maeztu”, a distinction awarded by the Spanish State Research Agency for the period 01/01/2026-31/12/2031.

\appendix
\crefalias{section}{appendix}

\section{Scalar reduction and principal symbol}
\label{app:principal_symbol}

For a homogeneous plasma with $\bm{B}_0=B_0\mathbf{\hat{e}}_z$, the vector wave equation in a two-dimensional $(z,x)$ domain is governed by the matrix differential operator
\begin{align}
	\mathcal{D} &= \nabla\times\nabla\times \:- \: k_0^2\bm{\kappa}, \qquad \mathcal{D}\bm{\hat{E}}=\bm{0}.
\end{align}
For constant plasma coefficients, computing the determinant of $\mathcal{D}$ gives a reduced scalar equation satisfied by each field component, $\det(\mathcal{D})\hat{E}_j=0$, with $j\in\{x,y,z\}$.
Expanding the determinant yields for fields independent of $y$
\begin{align}
	\det(\mathcal{D}) &= k_0^4 D^2(\partial_x^2+k_0^2P)-(\partial_x^2+\partial_z^2+k_0^2S)\big[(\partial_z^2+k_0^2S)(\partial_x^2+k_0^2P)-(\partial_x\partial_z)^2\big]. 
    \label{eqt:det_expansion}
\end{align}
Expanding the terms inside the square brackets cancels the fourth-order cross-derivatives. Defining the scalar operator $\mathcal{Q}=-k_0^{-2}\det(\mathcal{D})$, so that $\mathcal{Q}\hat{E}_j=0$, gives
\begin{align}
	\mathcal{Q} &= (\partial_x^2+\partial_z^2+k_0^2S)(P\partial_z^2+S\partial_x^2+k_0^2SP)-k_0^2D^2(\partial_x^2+k_0^2P).
\end{align}
Retaining only the fourth-order derivatives gives the principal part $\mathcal{Q}_4$ \cite{tayl91}, which factors into two second-order operators:
\begin{align}
	\mathcal{Q} &= (\partial_x^2+\partial_z^2)(S\partial_x^2+P\partial_z^2)+\text{l.o.t.} = \mathcal{Q}_4+\text{l.o.t.} 
    \label{eqt:principal_part}
\end{align}
Here, the lower-order terms contain at most second-order spatial derivatives. Substituting $\partial_x\rightarrow i\xi_x$ and $\partial_z\rightarrow i\xi_z$ in the principal part gives the symbol in \cref{eqt:principal_eqt}.

%%%%%%%%%%%%%%%%%%%%%%%%%%%%%%%%%%%%%%%%%%%%%%%%%%%%%%%%

\section{Plane-wave residual analysis} 
\label{subsec3_NUM_DISP}
Classical FE dispersion analyses determine the discrete Bloch modes supported by periodic meshes, allowing independent amplitudes for the distinct DoFs within the unit cell. Such analyses have been developed for Maxwell discretizations on triangular grids \cite{monk1994} and high-order Nédélec elements on tensor-product grids \cite{ainsworth2004}.

Here, we adopt an interpolated plane-wave residual analysis based on the FE operators assembled through \texttt{Gridap.jl}. A continuous plane wave is interpolated into the mixed FE space, linking its discrete DoFs through the three Cartesian field amplitudes. This choice is motivated by the plane-wave fields of the analytical Fresnel reference and provides a diagnostic of how the assembled operator responds to these fields. This analysis complements the analytical and FFT diagnostics, but does not cover the full discrete Bloch spectrum.
 
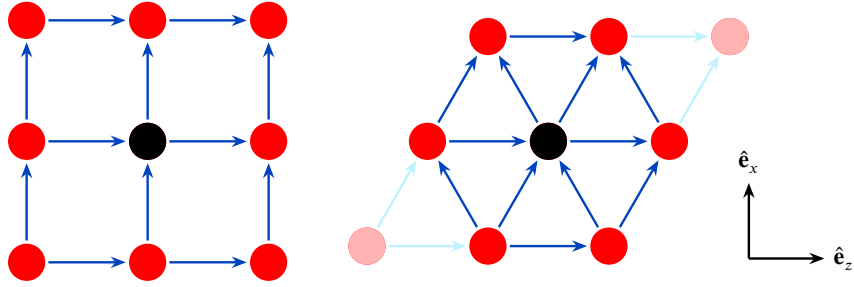
\begin{figure}[pos=H]
    \centering
    \begin{tikzpicture}[
    x=1cm, y=1cm,
    vertex/.style={circle, fill=red, draw=none, inner sep=0pt, minimum size=14pt},
    reference/.style={vertex, fill=black},
    faded vertex/.style={vertex, fill=red!30},
    edge/.style={-{Stealth[length=2mm,width=1.5mm]}, draw=blue!65!cyan!75!black, line width=0.9pt, shorten <=1pt, shorten >=1pt},
    faded edge/.style={edge, draw=blue!20!cyan!25},
    axis/.style={-{Stealth[length=2mm,width=1.5mm]}, line width=0.9pt}
]

    \def\h{1.6}

    % Rectangular mesh: a 2 x 2 patch.
    \foreach \i in {0,1,2} {
        \foreach \j in {0,1,2} {
            \node[vertex] (R\i\j) at ({\h*\i},{\h*\j}) {};
        }
    }
    \node[reference] at (R11) {};

    \foreach \j in {0,1,2} {
        \draw[edge] (R0\j) -- (R1\j);
        \draw[edge] (R1\j) -- (R2\j);
    }
    \foreach \i in {0,1,2} {
        \draw[edge] (R\i0) -- (R\i1);
        \draw[edge] (R\i1) -- (R\i2);
    }

    % Triangular mesh: the split 2 x 2 patch mapped to an equilateral lattice.
    \begin{scope}[shift={(4.5,0.21436)}]
        \foreach \i in {0,1,2} {
            \foreach \j in {0,1,2} {
                \node[vertex] (T\i\j) at ({\h*(\i+\j/2)},{\h*sqrt(3)*\j/2}) {};
            }
        }
        \node[reference] at (T11) {};
        \node[faded vertex] at (T00) {};
        \node[faded vertex] at (T22) {};

        % Horizontal edges.
        \draw[faded edge] (T00) -- (T10);
        \draw[edge] (T10) -- (T20);
        \draw[edge] (T01) -- (T11);
        \draw[edge] (T11) -- (T21);
        \draw[edge] (T02) -- (T12);
        \draw[faded edge] (T12) -- (T22);

        % Edges oriented upwards and to the right.
        \draw[faded edge] (T00) -- (T01);
        \draw[edge] (T01) -- (T02);
        \draw[edge] (T10) -- (T11);
        \draw[edge] (T11) -- (T12);
        \draw[edge] (T20) -- (T21);
        \draw[faded edge] (T21) -- (T22);

        % Edges oriented upwards and to the left.
        \draw[edge] (T10) -- (T01);
        \draw[edge] (T20) -- (T11);
        \draw[edge] (T11) -- (T02);
        \draw[edge] (T21) -- (T12);
    \end{scope}

    \begin{scope}[shift={(9.55,0.05)}]
        \draw[axis] (0,0) -- (1,0) node[right] {$\hat{\mathbf{e}}_z$};
        \draw[axis] (0,0) -- (0,1) node[above] {$\hat{\mathbf{e}}_x$};
    \end{scope}

\end{tikzpicture}
    \caption{Rectangular (left) and triangular (right) mesh patches used in \texttt{Gridap.jl} to assemble the discrete operators for the plane-wave residual analysis. The central reference node is shown in black, the nodes of the surrounding elements in red, and their oriented edges in blue.}
    \label{fig:3_mesh_Dofs}
\end{figure}

The analysis is performed on a planar periodic mesh in order to isolate the intrinsic dispersive properties of the discretization.
Two structured reference mesh families are considered: quadrilateral meshes composed of square elements, and triangular meshes arranged in a regular hexagonal lattice (\cref{fig:3_mesh_Dofs}).

The spatial distribution of the DoFs within these meshes determines the size of the reduced plane-wave residual matrix.
For first-order discretizations, each vertex carries one nodal DoF and each edge carries one tangential DoF. Second-order approximation introduces additional nodal, edge, and interior DoFs. 
The coexistence of nodal and edge-based unknowns, combined with the anisotropic dielectric tensor $\bm{\kappa}$, results in a fully coupled discrete system across all field components.
The plane-wave residual analysis combines the assembled FE operators together with the interpolation of a continuous plane wave into the corresponding discrete space. 
It can therefore be applied to different element types and polynomial orders without a separate analytical derivation for each case.

Let $N_\mathrm{DoFs}$ denote the total number of DoFs in the assembled mesh, and $\bm{U} \in \mathbb{C}^N_\mathrm{DoFs}$ the global vector of discrete DoFs resulting from the FE discretization. The assembled system associated with the weak formulation in a planar, homogeneous problem can be written as
\begin{equation}
   \sum_{j=1}^{N_\mathrm{DoFs}} \mathrm{D}_{ij} U_j = 0, \quad \text{for } i = 1 \dots N_\mathrm{DoFs},
\end{equation}
where $\mathbf{D} = \bm{K} - k_0^2 \bm{M}$ is the dynamic matrix, with $\bm{K}$ the curl–curl stiffness matrix and $\bm{M}$ the dielectric-weighted mass matrix. Both matrices are assembled directly from the weak formulation using \texttt{Gridap.jl}.
The indices $i$ and $j$ range from $1$ to $N_\mathrm{DoFs}$, such that the global system matrix $\mathbf{D}$ is an $N_\mathrm{DoFs} \times N_\mathrm{DoFs}$ matrix.

To evaluate the discrete response to a continuous plane wave, we impose a discrete plane wave ansatz parameterized by a wavevector $\bm{k}$.
The corresponding electric field phasor is written as
\begin{equation}
    \bm{\hat{E}}(\bm{r}) = \bm{\tilde{E}}\exp(i\bm{k}\cdot\bm{r}), \qquad \bm{\tilde{E}} = (\tilde{E}_x,\tilde{E}_y,\tilde{E}_z)^{\mathsf{T}},
\end{equation}
where $\bm{\tilde{E}}$ is the constant complex amplitude vector, containing the magnitudes and relative phases of the field components.
At a reference location $\bm{r}_s$ in the mesh, the spatial exponential factor is denoted by
\begin{equation}
    \varphi_s(\bm{k}) = \exp(i\bm{k}\cdot\bm{r}_s).
\end{equation}
The corresponding discrete DoFs are obtained by interpolating this continuous plane wave phasor into the mixed FE space.  In practice, this operator is constructed via the interpolation routines of the FE framework.
The discrete field is thus written as:
\begin{equation}
    U_j = \sum_{s=1}^{N}\sum_{l=1}^{3}\mathcal{I}_{jsl}\tilde{E}_l\varphi_s(\bm{k}), \qquad j=1 \ldots N_\mathrm{DoFs},
\end{equation}
where the index $s$ runs over the $N$ spatial support points associated with the DoFs, and $l$ indexes the three Cartesian components.
The coefficients $\mathcal{I}_{jsl}$ account for the geometric mappings, DoF orientations, and interpolation weights.
Substituting this ansatz into the global dynamic system generates a residual for each discrete test equation:
\begin{equation}
    (\mathbf{D}\bm{U})_i = \sum_{l=1}^3 R_{il}\tilde{E}_l , \qquad R_{il}(\bm{k}) = \sum_{j=1}^{N_\mathrm{DoFs}}\sum_{s=1}^{N}\mathrm{D}_{ij}\mathcal{I}_{jsl}\varphi_s(\bm{k}).
\end{equation}
Under the plane-wave ansatz, the same discrete equations
repeat from one mesh cell to the next, differing only by a spatial phase factor.
To characterize the local discrete response, we thus retain a set of $d$ equations around the central reference node, covering the distinct local DoF types and excluding equations affected by the mesh patch boundary.

Denoting the retained row indices by $i_1,\ldots,i_d$, the reduced residual matrix $\bm{\mathbf{R}}(\bm{k})\in\mathbb{C}^{d\times3}$ is defined by
\begin{equation}
    \mathbf{R}_{al}(\bm{k}) = R_{i_a l}(\bm{k}), \qquad a=1,\ldots,d.
\end{equation}
A nonzero amplitude vector satisfies this reduced system when $\bm{\mathbf{R}}$ loses column rank. We assess this condition through its singular value decomposition rather than a determinant. For $d\geq3$, this decomposition is:
\begin{equation}
    \bm{\mathbf{R}} = \bm{U}_{\mathrm{s}}\bm{\Sigma}\bm{V}_{\mathrm{s}}^\dagger, \qquad \bm{\Sigma}=\operatorname{diag}(\sigma_1,\sigma_2,\sigma_3), \qquad \sigma_1\geq\sigma_2\geq\sigma_3\geq0.
\end{equation}
Here, $\bm{U}_{\mathrm{s}}\in\mathbb{C}^{d\times3}$ and $\bm{V}_{\mathrm{s}}\in\mathbb{C}^{3\times3}$ contain the left and right singular vectors, respectively.
The smallest singular value measures the minimum residual over all unit-norm amplitude vectors:
\begin{equation}
    \sigma_{\min}(\bm{k})=\sigma_3(\bm{k})=\min_{\|\bm{\tilde{E}}\|_2=1}\|\bm{\mathbf{R}}(\bm{k})\bm{\tilde{E}}\|_2.
\end{equation}
Thus, $\sigma_{\min}=0$ indicates an exact nontrivial solution of the reduced system within the adopted ansatz, while small values identify interpolated plane waves with small discrete residuals.

\bibliographystyle{elsarticle-num}
\bibliography{bibtex/ep2,bibtex/others,bibtex/Zotero}

\end{document}